\documentclass[11pt,preprint]{aastex}
\usepackage{epsfig}

\begin{document}

\title{Studying the Galactic Bulge Through Spectroscopy of
Microlensed Sources:\\ II. Observations\altaffilmark{1}}

\author{Stephen R. Kane\altaffilmark{2} and Kailash C. Sahu}
\affil{Space Telescope Science Institute, 3700 San Martin Drive,
Baltimore, MD 21218, U.S.A.}
\email{skane@stsci.edu, ksahu@stsci.edu}

\altaffiltext{1}{Based on observations collected at the European Southern
Observatory, La Silla, Chile}
\altaffiltext{2}{Currently at the School of Physics \& Astronomy,
University of St Andrews, North Haugh, St Andrews, Fife  KY16 9SS,
Scotland}

\begin{abstract}

The spectroscopy of microlensed sources towards the Galactic bulge
provides a unique opportunity to study (i) the kinematics of the
Galactic bulge, particularly its far-side, (ii) the effects of
extinction on the microlensed sources, and (iii) the contributions
of the bulge and the disk lenses to the microlensing optical depth.
We present the results from such a spectroscopic study of 17
microlensed sources carried out using the ESO Faint Object
Spectrograph (EFOSC) at the 3.6 m European Southern Observatory (ESO)
telescope. The spectra of the unlensed sources and Kurucz model
spectra were used as templates to derive the radial velocities and the
extinctions of the microlensed sources. It is shown that there is an
extinction shift between the microlensed population and the
non-microlensed population but there is no apparent correlation between
the extinction and the radial velocity. This extinction offset, in our
best model, would imply that 65\% of the events are caused by
self-lensing within the bulge. The sample needs to be increased to
about 100 sources to get a clear picture of the kinematics of the
bulge.

\end{abstract}


\section{INTRODUCTION}

By now, several authors \citep{kir94,pac94b,mol96,zha95} have pointed
out that the stars within the Galactic bulge may play a dominant role as
gravitational lenses, and a significant fraction of the detected events
may be due to lensing by stars within the bulge. If a significant
fraction of these events are indeed due to lenses within the bulge, then
the microlensing characteristics can provide a powerful technique
of deriving the stellar mass density and mass function within the
Galactic bulge \citep{zha95}. However, the contributions of different
populations to the microlensing optical depth remain uncertain, and their
determination is important in using the microlensing characteristics to
derive other quantities such as the stellar mass density and the stellar
mass function.

If the lensing is caused predominantly by bulge stars, then a major
fraction of the lensed stars will be at the far side of the bulge so that
there are enough stars in front to cause the lensing. Thus they would be
fainter in general, and it was shown by \citet{sta95} that the magnitude
offset between the lensed sources and non-lensed sources can be a good
measure of the fraction of the events caused by bulge lenses. A similar
test was suggested by \citet{kan00b} (hereafter referred to as paper I)
through the measurement of extinction and it was shown that an extinction
offset may be measurable from the spectra of lensed and non-lensed
sources. This test used the principle that the lensed stars should have
larger extinction since they would predominantly be on the far side.

If the sources are predominantly in the far side of the bulge, then the
spectra of these sources give a unique opportunity to derive the radial
velocities of the objects in the far side of the Galactic bulge. The
radial velocities derived from the observed spectra can be combined with
the proper motion derived from the microlensing time scales to determine
the 3-dimensional velocity structure of the far side of the Galactic
bulge.

This paper presents the results of a spectroscopic study of microlensed
sources conducted using data obtained with the 3.6 m telescope of the
European Southern Observatory (ESO) at La Silla, Chile. Kurucz model
spectra were used to create theoretical extinction effects for various
spectral classes of stars towards the Galactic bulge. These extinction
effects are used to interpret spectroscopic data consisting of a sample
of microlensed stars towards the Galactic bulge. The extinction offsets
of the lensed source with respect to the average population is derived
and a measurement of the fraction of bulge-bulge lensing is made.
Measurements of the radial velocities of these sources are used as an
attempt to determine the kinematic properties of the far side of the
Galactic bulge.

\section{OBSERVATIONS AND DATA REDUCTION}

\subsection{Details of the Observations}

The observations were taken with the ESO 3.6 m telescope using 
the ESO Faint Object Spectrograph and Camera (EFOSC). The data were
obtained in two observing runs: the first run was from June 28 to July 1,
1995 during which a major part of the observations were taken, and the
second run was on the night of June 14, 1996 during which only one
source was observed. The sky condition was clear during the whole 1995
observing run; the 1996 observations, however, were affected by passing
cirrus and clouds. Table 1 shows the grisms that were used for the
observations. The CCD detector used has $512 \times 512$
pixels and a pixel scale of 0.61 arcsec pixel$^{-1}$ both in the
dispersion and the cross-dispersion directions. The seeing during the
observations was $\sim$1 arcsec, and a slit-width of $1.5''$ was used
for all the observations. The spectral resolution depends on the chosen
slit width, and for our observations, the true spectral resolution is
approximately 2.46 times the dispersion (\AA\ per pixel) shown in Table
1. At the time of observations, the camera was set to have a gain of 3.8
electrons per count for which the read-out noise is 8.5 electrons per
pixel.

As the name implies, the EFOSC is both an imaging and a spectrographic
camera. The imaging capability of this spectrograph greatly facilitates
the spectroscopic observations of stars in crowded fields such as the
Galactic bulge by examining the field first, before placing the slit on
the object of interest. This facility was particularly useful for our
purpose of simultaneously obtaining the spectra of a few
non-microlensed sources along with the microlensed source, by placing
them all along the same slit. In order to achieve this, an image was
obtained first, with a typical integration time of about 2 minutes.
This image was taken using a clear (unfiltered) aperture to avoid any
shift while placing the slit. Then the slit was centered on the
microlensed source and its orientation was set such that (i) it was
close to the East-West direction (so that the differential atmospheric
dispersion was minimum along the slit), and (ii) there were a
sufficient number of non-lensed sources on the slit (which would be
later used as the control sample in the analysis). Since the size of
the detector is rather small ($512 \times 512$ pixels), the spectral
coverage in each setting is accordingly small. So it was generally
necessary to take two separate spectra using two grisms in the blue and
the red, in order to cover the wavelength range of our interest. This
limited spectral coverage in a single grism setting had the advantage
that the effect of the differential atmospheric dispersion at various
wavelengths in a spectrum is minimal. Nevertheless, attempts were made
to do the observations in low air-masses to further minimize any
differential atmospheric dispersion at different wavelengths in a 
given setting.

To correct for the electronic noise associated with the detector
readout, 5 bias exposures were taken every evening, the median of which
was used for subtracting the bias level. To correct for pixel-to-pixel
sensitivity variations of the detector, flat-field exposures were taken
by observing a part of the dome illuminated with a tungsten lamp.
Typically, 3 flat-field exposures were taken for each grism in the
beginning of each night, the average of which was used during the data
reduction.

For wavelength calibration, spectra were taken by illuminating the slit
with a He lamp, followed by an Ar lamp in the same exposure. The
combination of He and Ar was necessary to get enough lines both in the
blue and the red region of the spectrum simultaneously. The rms
deviation of the residuals from the dispersion relations for each grism
are shown in Table 1. The lamp spectra were taken for each grism at the
beginning and end of every night. The typical wavelength shift during a
night was less than 0.1 pixels, which is not more than the rms deviation 
of the residuals from the dispersion relations.  

Since this study is mostly interested in the relative fluxes and line
strengths rather than their absolute values, observing a single
standard star per night was deemed sufficient. The standard star LTT
9239 was used for the 1995 observations and the standard star LTT 8702
was used for the 1996 observations. The uncertainty in absolute flux
calibration would normally be in the range of 10--20\% in clear sky
conditions and worse if the sky is not photometric, however the
relative fluxes will be more accurate.

The spectra of 17 different microlensed sources were obtained during
these observing runs. Since the purpose of this study is to compare the
spectra of microlensed sources with a control sample of other stars, 5
non-lensed stars were chosen from each field whose spectra were also
reduced. The microlensing events observed are summarized in Table 2,
where $t_E$ is the time taken by the lens to cross the Einstein ring
radius. This information has been extracted from the MACHO alerts 
(http://darkstar.astro.washington.edu/) and from the OGLE and DUO
publications \citep{ala95,woz98}. The binary events have an undetermined
characteristic time scale $t_E$.  The galactic longitude and latitude of
the sources lie in the range $0.55\degr<l<3.98\degr$ and
$-4.92\degr<b<-2.68\degr$ respectively. A summary of the observations of
these events is shown in Table 3. DUO 95-BLG-2 and OGLE 95-BLG-3 were
also observed, but the sky was scattered with clouds during these
observations. As a result, the signal-to-noise in these spectra are
considerably lower. Hence these two sources and their associated
non-microlensed stars are not included in the following analysis.

\subsection{Data Reduction}

After the usual bias subtraction, the wavelength calibration for the
entire 2-dimensional image was carried out through the lamp spectra
obtained with the appropriate grism during that particular night.
The 1-dimensional spectra were extracted for the microlensed source and
5 non-microlensed sources in the image using an extraction height of
5--7 pixels for each source. The images being crowded, sky subtraction
was tricky, so care was taken in choosing an appropriate region for the
sky subtraction. The spectra were corrected for atmospheric extinction
using a model available for the ESO observing site (for details, see
the MIDAS manual). The resulting 1-dimensional spectra were flux
calibrated through a standard star observation carried out in the same
night. The flat-field images obtained with the illuminated dome were
found to be relatively smooth, and the pixel-to-pixel response variation
was found to be small, so the flat-field images were not used in the
reduction. Instead, the flat-fielding and the flux calibration was done
through a single step through the standard star observations. These
final 1-dimensional spectra are then used for further analysis.

\section{ESTIMATING SPECTRAL TYPE, EXTINCTION, AND RADIAL VELOCITY}

In order to process the large number of spectra that were obtained, a
MIDAS script was written which estimates both the spectral type and the
extinction for each individual spectrum. This script uses a large library
of model spectra that was constructed from the Kurucz database and a
cross-correlation technique to achieve this, as explained in more detail
later. An additional script was written to determine the radial
velocities of the microlensed and non-microlensed sources which uses a
high signal-to-noise template spectrum and a cross-correlation technique.
More details of the method used in these scripts are given in Section 3.2.
(The script itself, with ample comments describing the algorithms, is
published by \citet{kan00a}).

\subsection{The Model Spectra}

The 1993 Kurucz stellar atmospheres atlas \citep{kur93} covers a wide
range of metallicities, effective temperatures, and gravities. The
models were first developed by Kurucz in 1970 using the stellar
atmosphere modeling program ATLAS \citep{kur70}. The 1993 atlas
contains about 7600 models which are convenient to access using the
IRAF (Image Reduction Analysis Facility) task {\em synphot}, which is
available in the {\em stsdas} package developed at the Space Telescope
Science Institute. The task {\it synphot} was used to extract a full grid
of model spectra from the atlas, by specifying the appropriate range of
temperatures, metallicities and the surface gravities, as explained below. 

{\begin{figure}
\plotone{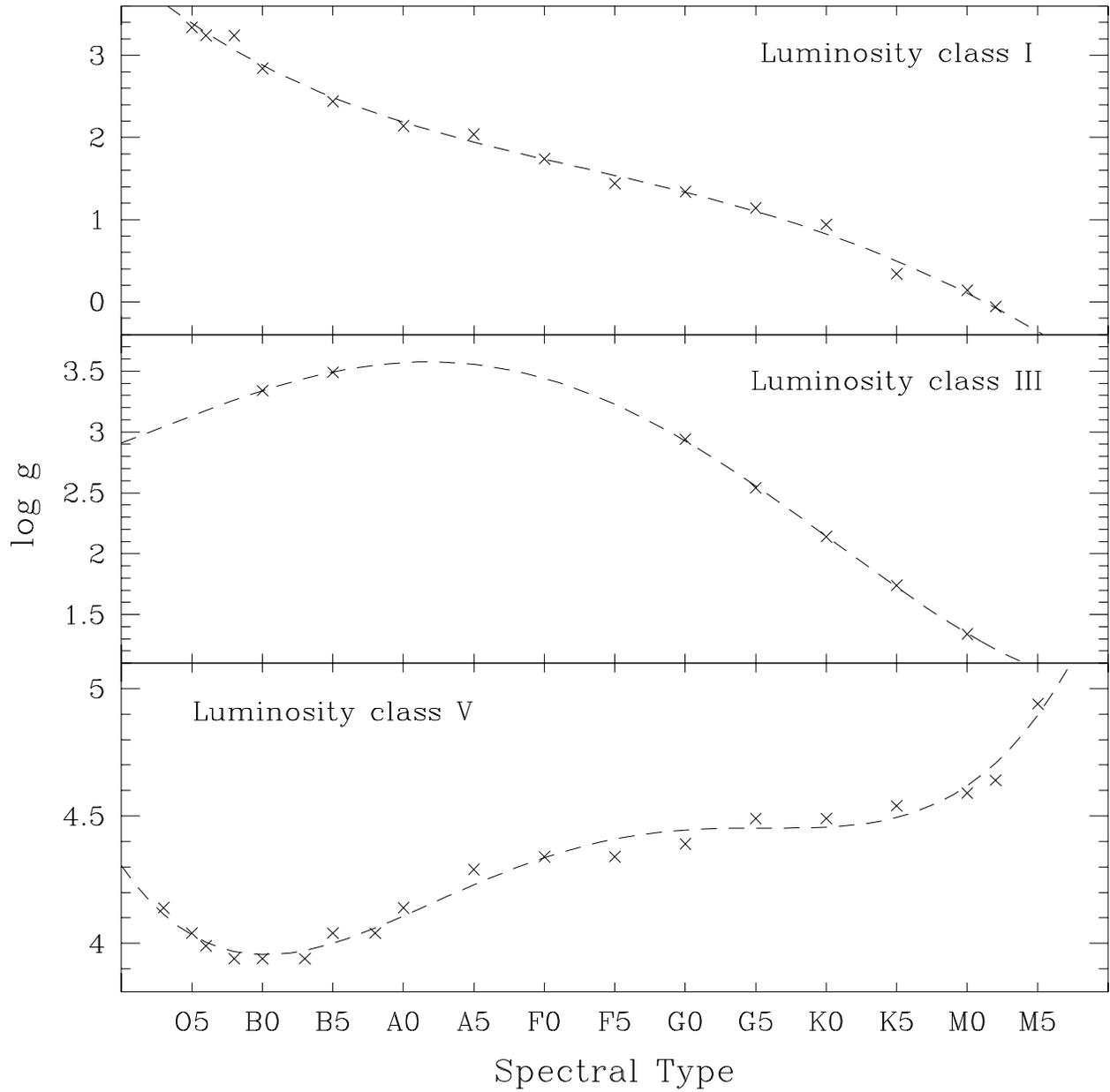}
\figcaption{Surface gravity $\log g$ for luminosity classes I, III,
and V.}
\end{figure}}

To extract the model spectra from the atlas, an exhaustive list of
stellar parameters was needed for each of the stellar sub-types and
luminosity classes. The values for the effective temperature $T_{eff}$,
the surface gravity $\log g$, and the absolute magnitude $M_V$
characterizing each star were obtained from Schmidt-Kaler's compilation
of physical parameters of stars \citep{sch82}. However, the values of
$\log g$ were found to be severely lacking in their coverage of the
stellar sub-types and so the remaining values were determined from the
interpolation of the values given by Schmidt-Kaler. As shown in Figure
1, a fourth-order polynomial of the form $y = ax^4+bx^3+cx^2+dx+e$ was
fitted to the available data for each luminosity class. The
coefficients for the polynomial fitted to each luminosity class are
shown in Table 4. The estimates obtained for the required values of
$\log g$ by this method were sufficient to create the desired spectral
models since spectral classification within a luminosity class has a
weak dependence on surface gravity. It is worth mentioning that the
grids of theoretical isochrones calculated by \citet{ber94} also provide
a useful set of stellar parameters. Although there are slight
differences in the model parameters between these two references, they
are too small to affect the main results of this study. A total of 226
spectra were produced for this analysis. Galactic bulge stars tend to
have a large range of metallicity. Although most stars are thought to
be more metal poor than solar \citep{see99}, recent work suggests that
there is a large dispersion in the stellar metallicity in the Galactic
bulge stars, ranging from 0.2 solar to more than solar \citep{fel00,sta00}.
We used an extensive library of model spectra which included
metallicity ranges from high metallicity of approximately solar to
low-metallicity of 0.1 times solar corresponding to population II stars.

\subsection{Fitting Routine for Determining Spectral Class and Extinction}

Spectral classification is based on the strength of various spectral
features in the spectra which are compared with those of a set of
standard stars defining the classification system. Although the
classification of the observed spectra is not the primary goal of this
analysis, it serves as a tool in deriving the necessary information
such as the extinction and the radial velocity.

The main purpose of this MIDAS fitting routine was to provide a
reasonable estimate of the extinction present in each of the measured
spectra. The fitting routine first combines spectra of a star observed
through multiple grisms into a single spectrum in order to increase the
reliability of the spectral classification. The fitting routine then
sequentially compares the spectrum with each of the model spectra
from the constructed model spectral library. The details of this
routine will now be outlined.

The observed spectrum was divided by the model spectrum and then linear
regression was used to calculate the slope of the resulting image. 
This step was then repeated, incrementing the interstellar extinction
(Galactic extinction data from \citet{sea79}) applied to the model
spectrum with each repetition, until the slope was approximately equal
to zero. The value of $E_{B-V}$ used to achieve a slope of zero was
then adopted as the extinction value for that model spectrum.

Once the extinction was estimated for the model spectrum, the extincted
model was fitted to the observed spectrum. The flux of the model was
first approximated to match the flux of the observed spectrum by
normalizing the flux of the model to that of the observed spectrum. A
fitting parameter was calculated by measuring the mean flux for each 10
pixels along the entire wavelength range for both the observed spectrum
and the model. The squares of the differences of these two values were
added and this sum was taken as the fit parameter for that spectral
model, similar to that which is obtained in a $\chi^2$ analysis. This
method takes both the continuum as well as the spectral lines into 
account, but places more weight on the spectral lines in the spectral
classification. Thus, the model spectrum with the lowest value of the
fit parameter was the one that fitted the observed spectrum best, and
the corresponding spectral class and the extinction value was assigned
to this particular source.

It should be noted that, although the spectral lines are used in the
classification, the effect of any single spectral line (such as the NaD
feature) in estimating the extinction is minimal. The use of both the
shape of the spectrum as well as the spectral lines considerably reduces
the degeneracy between the stellar spectral type and the extinction. 

{\begin{figure}
\plotone{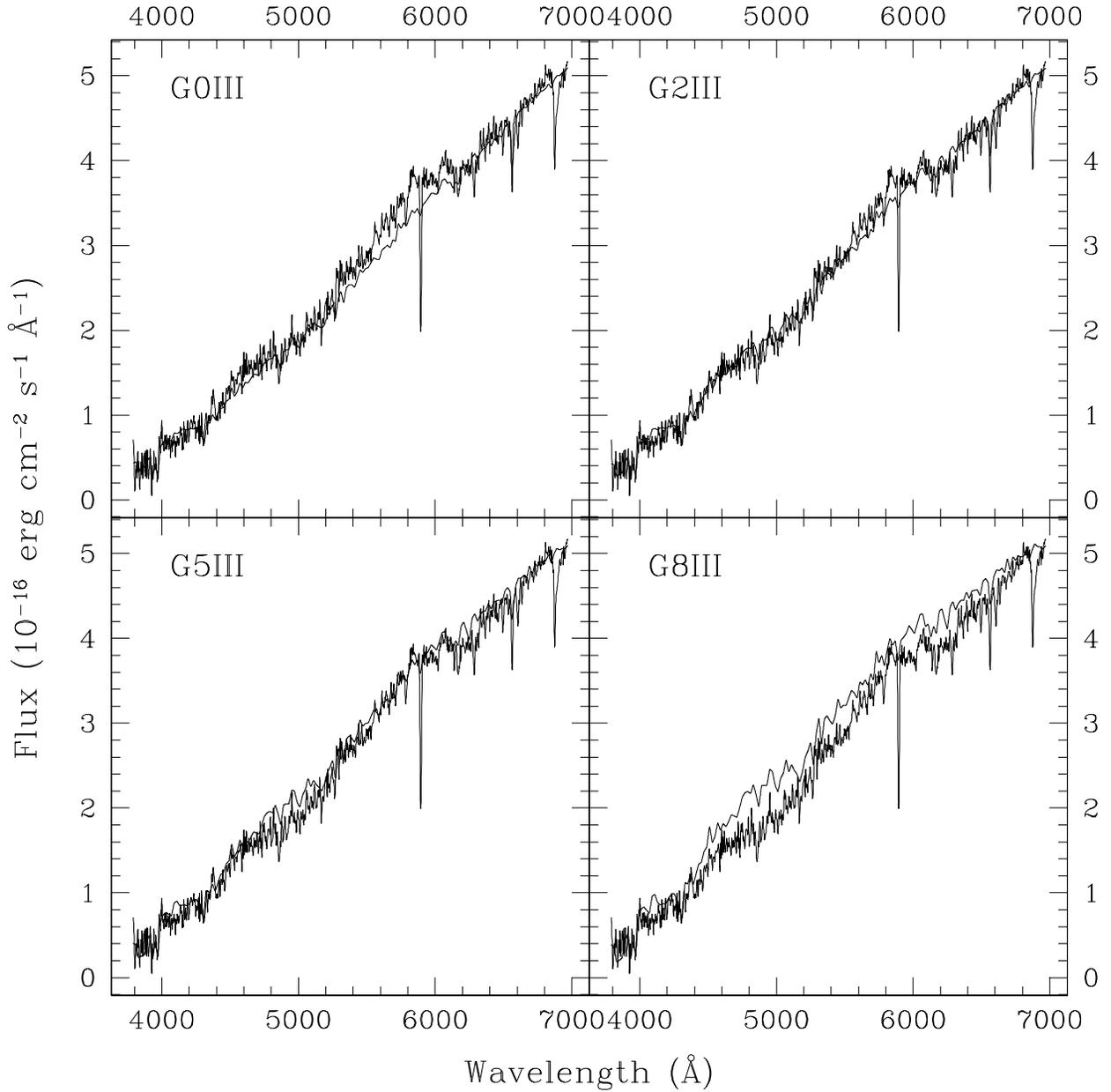}
\figcaption{Event MACHO 95-BLG-10 fitted with four different Kurucz
models. The fitted extinction values for the models G0III, G2III, G5III,
and G8III are 1.1, 0.99, 0.86, and 0.71 respectively. The best fitting
model, G2III, is shown in the top-right panel.}
\end{figure}}

This is illustrated in Figure 2 where we show an example of various model
fits to the spectrum of MACHO 95-BLG-10. The fitted extinction values for
the models G0III, G2III, G5III, and G8III are 1.1, 0.99, 0.86, and 0.71
respectively. In this case the G2III model provides the best fit to the
data. Adding larger extinction to earlier-type models or smaller extinction
to later-type models quickly deteriorates the fit. This implies that the
value of extinction derived from this method is good to within $\pm$0.1
for individual stars. 

Clearly, the error in the extinction is not necessarily a reflection of
the error in the fit but rather it is a reflection of the difference in
fitted extinction between the models. Several steps were taken to make
sure that this method works correctly, which are described below. 

First, to guard against any possible flaws in the fitting routine, the
spectral classification of stars using the fitting routine was
accompanied by a manual comparison with standard stars from libraries of
stellar spectra \citep{dan94,jac84,tor93}. The MK classification of
stellar spectra \citep{mor43} provides a complete and detailed
2-dimensional classification system for the classification of stars of
almost the entire spectral sequence. This method consists of estimating
relative intensities of suitable lines, some of which are sensitive to
temperature and some others sensitive to luminosity. The luminosity
and temperature dependence of many of the main spectral lines have been
conveniently summarized by \citet{tor93} which was used to cross-check
the classifications. 

We note that our fitting routine described here uses a combination of
line ratios and the shape of the entire continuum. So our
classification should be at least as reliable as the MK classification
method. As an example, the method described here provides a
classification of M2III for MACHO 95-BLG-30, a close match (within the
uncertainties) to the result of M4III derived by the MACHO
collaboration \citep{alc97} (More on this comparison in Section 4). 

Second, we have also used the luminosities as a safeguard against the
`degenerate' models. The observed luminosities are expected to be
different for different spectral type, and this information can be used
to choose the correct model. As explained later, the observed luminosities
are consistent with that expected from the model which suggests that the
derived models are reasonable.    

The limitations in the model spectra also play some role in the spectral
classification. The model spectra cover a wavelength range from the
ultra-violet (1000 \AA) to the infra-red (10 $\mu$m). However, the
model spectra are particularly unreliable for wavelengths greater
than 9000 \AA, largely due to very strong atmospheric water band
extinction, and indeed spectral information in this region has only
been obtained in recent years (after the models were created). To
account for this, wavelengths greater than 9200 \AA \ were ignored for
spectra obtained through the R300 grism. However, R300 observations are
available only for one source (MACHO 95-BLG-13). Furthermore, this
source has been observed both in B150 and O150 gratings with good 
signal-to-noise making the spectral classification fairly secure.

There is a truncation error in the stellar parameters used for each
model due to the limitations in the grid of models available from the
Kurucz stellar atmospheres atlas. This resulted in a limitation in the
number of models that could be produced and consequently in the
resolution of the fitting procedure. The lower threshold in the grid of
temperatures of 3500 K meant that stars cooler than spectral types of
about M2 could not be created. This consideration led to an estimated
uncertainty in the classification of about two spectral subtypes if it
is M2 or later. However, we do not expect any of the observed stars to
be later than M2 (because of the magnitude limit of the sample), and
hence this limitation in unlikely to have a significant effect on our
analysis.

It should be noted that for 8 of the 17 observed events, spectra were 
only obtained through the O150 grism. The limited wavelength range in
these spectra (5230--6970 \AA) made the fitting of an adequate model a
more challenging task. Further limitations were found when the fitting
routine attempted to fit models for late K and early M-type stars,
particularly for stars which were only observed using the O150 grism.
The spectral bands (such as the TiO band that is characteristic of
M-type stars) that tend to dominate these stellar types created problems
which in some cases caused a mis-classification of the spectra. In these
cases, special care was taken to identify the spectra through the use of
the previously mentioned libraries of stellar spectra and their
luminosities.

It is important to remember, however, that the exact spectral
classification is not important for our purpose as long as the
extinction values are not severely affected. As explained above, we
have taken several precautionary measures in estimating the spectral
classes and extinction values. The error in the extinction of an
individual star may be high, but we are interested in comparing the
extinctions of a sample as a whole. Hence the derived extinction values
should be adequate for the statistical investigation that we intend to
undertake in this study.

\subsection{Procedure for Radial Velocity Determination}

In general, the radial velocity of a star may be measured from the
Doppler shift of stellar spectral lines. The radial velocity is then
given by $v_r = (\Delta \lambda / \lambda_0) c$ where $\Delta \lambda
= \lambda - \lambda_0$ is the Doppler shift of the line from its rest
wavelength $\lambda_0$. However, there are factors intrinsic to stellar
structure, such as surface convection and magnetic fields, which can
affect the symmetry and wavelength of line profiles \citep{dra99}. An
approximate value of the radial velocity may still be determined from
one of the few lines, such as H$\alpha$ (6563 \AA), which are less
sensitive to the velocity structure of the photosphere.

A more reliable and accurate method for measuring the radial velocity
of a star is to cross-correlate the stellar spectrum with a template
spectrum. The correlation between them may be analyzed using the
cross-correlation function from which the location of the main peak is
used to determine the wavelength shift. Cross-correlation techniques
and the theory of correlation analysis have been described in detail by,
for example, \citet{ton79}.

To obtain absolute radial velocities (radial velocities relative to the
barycenter of the solar system), it is often necessary to
cross-correlate the stellar spectrum with that of a radial velocity
standard star, such as those monitored by CORAVEL \citep{udr99}. No
such standard stars were observed to carry out such an analysis since
we are mostly interested in the relative radial velocities. As noticed
earlier by \citet{mor91}, when a large number of spectral lines are used
for the radial velocity determination, the systematic errors caused by
lines formed at different regions of the stellar atmosphere average out,
and the resultant radial velocity determination is insensitive to the
choice of template for late-type stars. So the radial velocities
measured relative to a template bright star are adequate for this
analysis. The template used for these measurements (see Figure 3) was a
bright star with high S/N selected from the MACHO 95-BLG-12 field. By
fitting a gaussian to the H$\alpha$ line, the absolute radial velocity
of the template star was found to be $-98.5 \pm 18.0 \, \mathrm{km \,
s^{-1}}$. Taking into account the wavelength calibration residuals, the
absolute radial velocity is correctly stated as $-98.5 \pm 37.3 \,
\mathrm{km \, s^{-1}}$. Note that the random errors in the data, which
are subsequently used for the cross-correlation, can significantly affect
the results. So it is important to make sure that the signal-to-noise
(S/N) of the template spectrum is large (so that the effect of the random
noise is small). Our choice of a bright star as a template, and the
consequent high S/N of the spectrum, should help in this regard. It is
worth emphasizing here that we are mainly interested in the differential
radial velocities between the lenses and the sources. So the
uncertainties in the absolute velocity used for this star and the
associated uncertainties do not impact our analysis.

{\begin{figure*}
\epsfig{file=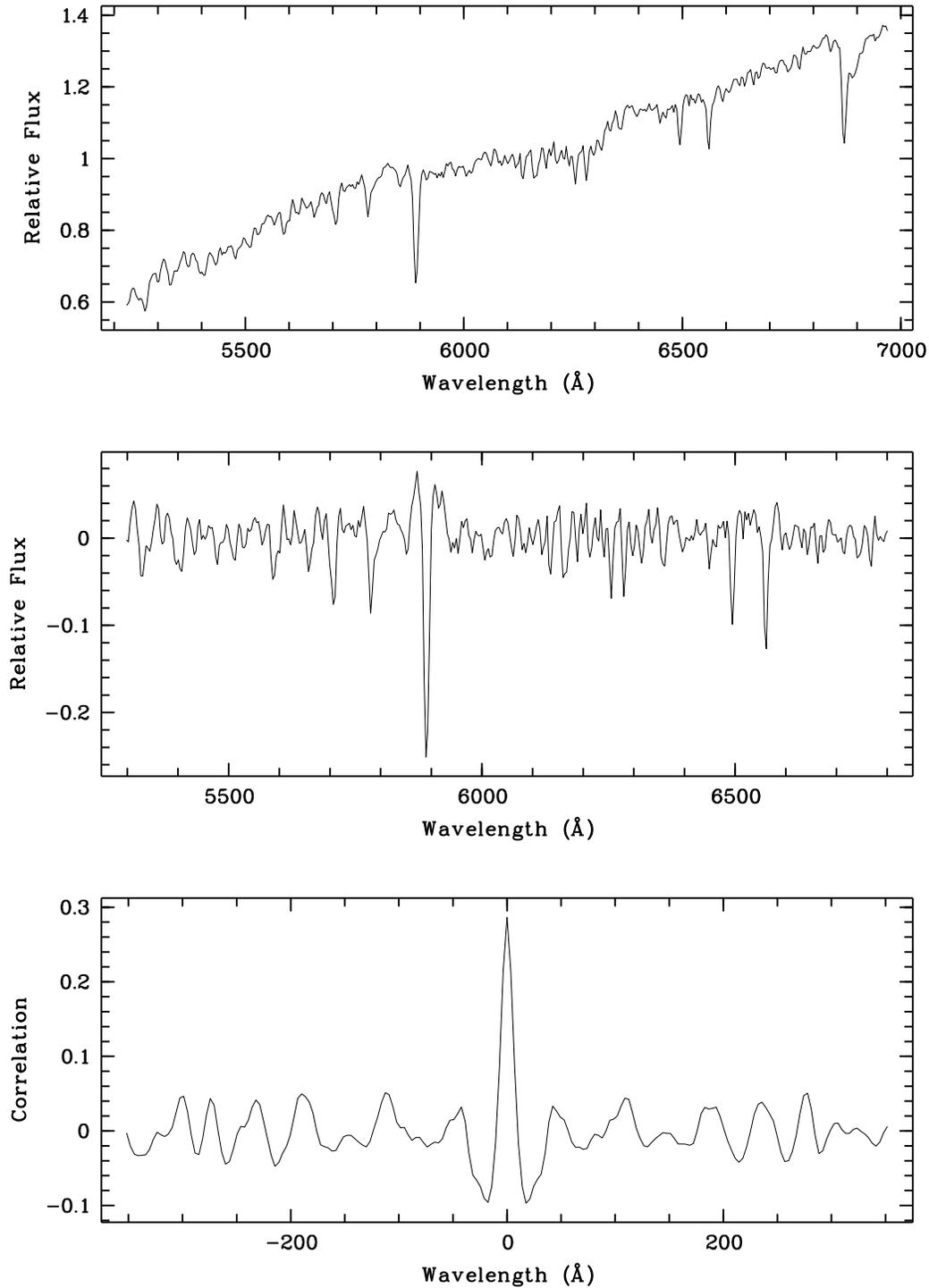, height=7.5in}
\caption{{\bf Top:} Normalized spectrum of the star to be used
as the cross-correlation template. {\bf Middle:} The result of subtracting
the continuum, applying a bandpass filter, and extracting the region of
the spectrum that excludes atmospheric lines. {\bf Bottom:} The
cross-correlation function that results from correlating the template
with itself. The shift of the central peak is approximately zero as
expected.}
\end{figure*}}

A MIDAS script was written to perform the cross-correlation and extract
the radial velocity information. Prior to performing the correlation, 
it is important to appropriately prepare a spectrum to reduce noise in the 
cross-correlation function. The first step in this process is to normalize
the spectrum. Next, an eighth-order polynomial is fitted to the continuum
and the continuum is subtracted. A bandpass filter is applied to the
spectrum which excises both high and low spatial frequency components.
The final step is to extract a large segment of the spectrum which
excludes atmospheric absorption lines. These steps are shown in Figure 3
in which the template spectrum is cross-correlated with itself.

After cross-correlating the stellar spectrum with the template spectrum,
the wavelength shift was determined from the position of the peak of the
cross-correlation function. As amply demonstrated by the recent extra-solar
planet detections through radial velocity measurements (see, eg.,
\citet{but96}), the uncertainty in the radial velocity measurements through
a cross-correlation technique can be substantially smaller than the
spectral resolution, if the S/N in the spectrum is large and the choice of
the template spectrum is appropriate. The central position of the
cross-correlation peak was determined by fitting a gaussian which has an
associated error. In order to determine the error in the cross-correlation
process, it was necessary to take a cross-correlation for which the result
was known and then add noise to one of the spectra. To do this, a program
was written which takes a spectrum of reasonable S/N and cross-correlates
the spectrum with itself several hundred times. With each iteration,
various wavelength shifts and noise levels were simulated in the duplicate
spectrum. The average difference between the calculated wavelength shift
and the real wavelength shift was used to estimate the error in the
cross-correlation process for different values of S/N. The results showed
that the error in the cross-correlation is around 5 km/s and only becomes
significant for very low S/N. Hence the cross-correlation algorithm is
fairly robust, mostly due to the number of lines used. This error was
combined with the error in the gaussian fit to estimate the total error
in the radial velocity. The MIDAS cross-correlation script was also
successfully tested by simulating line shifts in various spectra and by
performing correlations on restricted wavelength ranges within a spectrum.

The template spectrum of the bright star was cross-correlated with the
spectra of the lensed and the unlensed sources in each image, and
their radial velocities were determined as outlined above.

\section{RESULTS AND DISCUSSION}

For each source an estimate of the spectral type, extinction, and
relative radial velocity is made as discussed above. It is important 
to remember that the measurements of the microlensed sources apply 
only to the source and not the lens itself. The contribution of the
lens is assumed to be small because the stellar mass function is biased
towards lower masses and hence the lens is likely to be less massive
and considerably fainter than the source \citep{mao98}. Low luminosity
stars do not contribute significantly as sources since these would not
normally be detected by the magnitude limited microlensing surveys. 

{\begin{figure*}
\epsfig{file=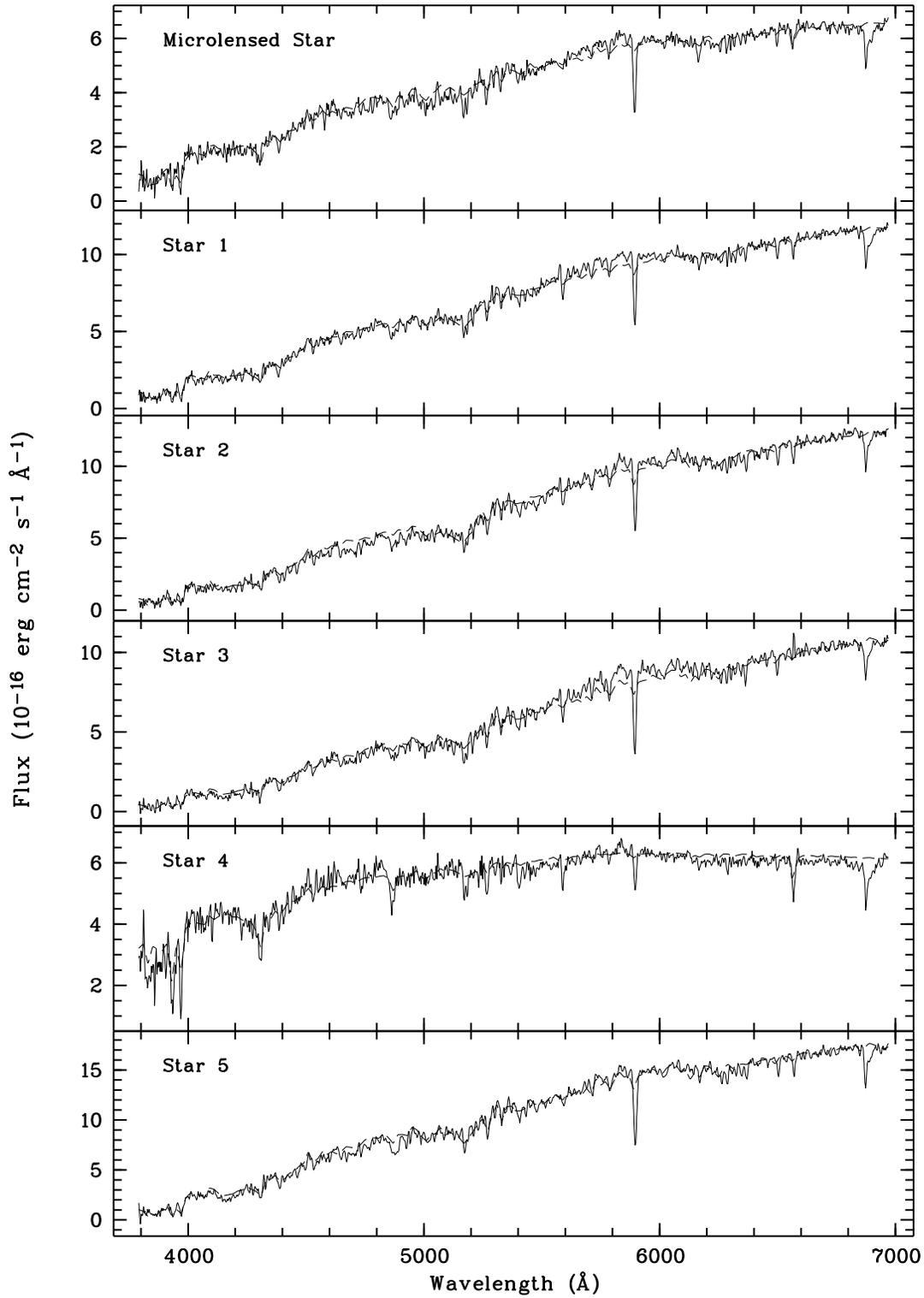, height=8in}
\caption{Spectra (solid line) and fitted models (dashed line)
for MACHO 95-BLG-17 and five non-microlensed stars in the field.}
\end{figure*}}

Table 3 gives the details of the observations, such as the grisms used, 
the date of observations, and the exposure times for all the
observations. Table 5 shows the results of the analysis for all the
observed microlensed sources. For the purpose of this table, the names
of the events have been abbreviated, the first letter indicating the
name of the collaboration and the following two numbers indicating the
identity of the event. The results include the estimated spectral class,
the color excess, and radial velocity, along with the uncertainties for
each source.

For illustration, more details of the analysis procedure are presented
for MACHO 95-BLG-17 which was observed with B150 and O150 grisms. For
this event, the observed spectra of the microlensed source and five
other stars in the field, along with their associated model spectra
are shown in Figure 4. The results from fitting models to the spectra
are given in Table 6. Shown in Figure 5 is the cross-correlation
function for the spectrum when cross-correlated with the chosen radial
velocity standard. 

{\begin{figure*}
\epsfig{file=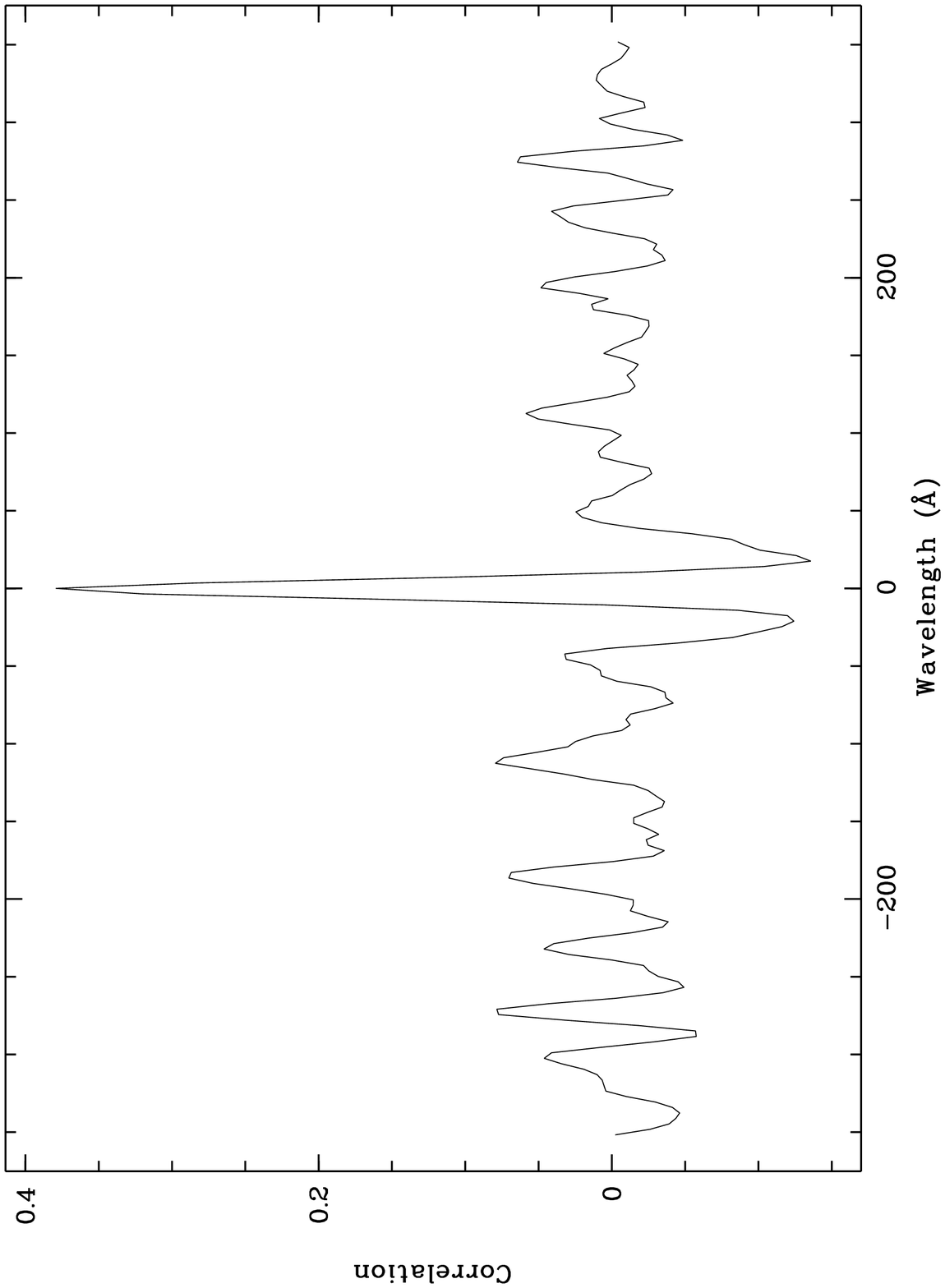,width=6in}
\caption{Cross-correlation function for the spectrum of
MACHO 95-BLG-17.}
\end{figure*}}

The spectra of the remainder of the microlensed sources are shown in
Figures 6--9. Note that most stars were fitted better by the lower 
metallicity models. This is an expected result since the Galactic bulge
tends to be dominated by population II stars \citep{see99}. 

{\begin{figure}
\plotone{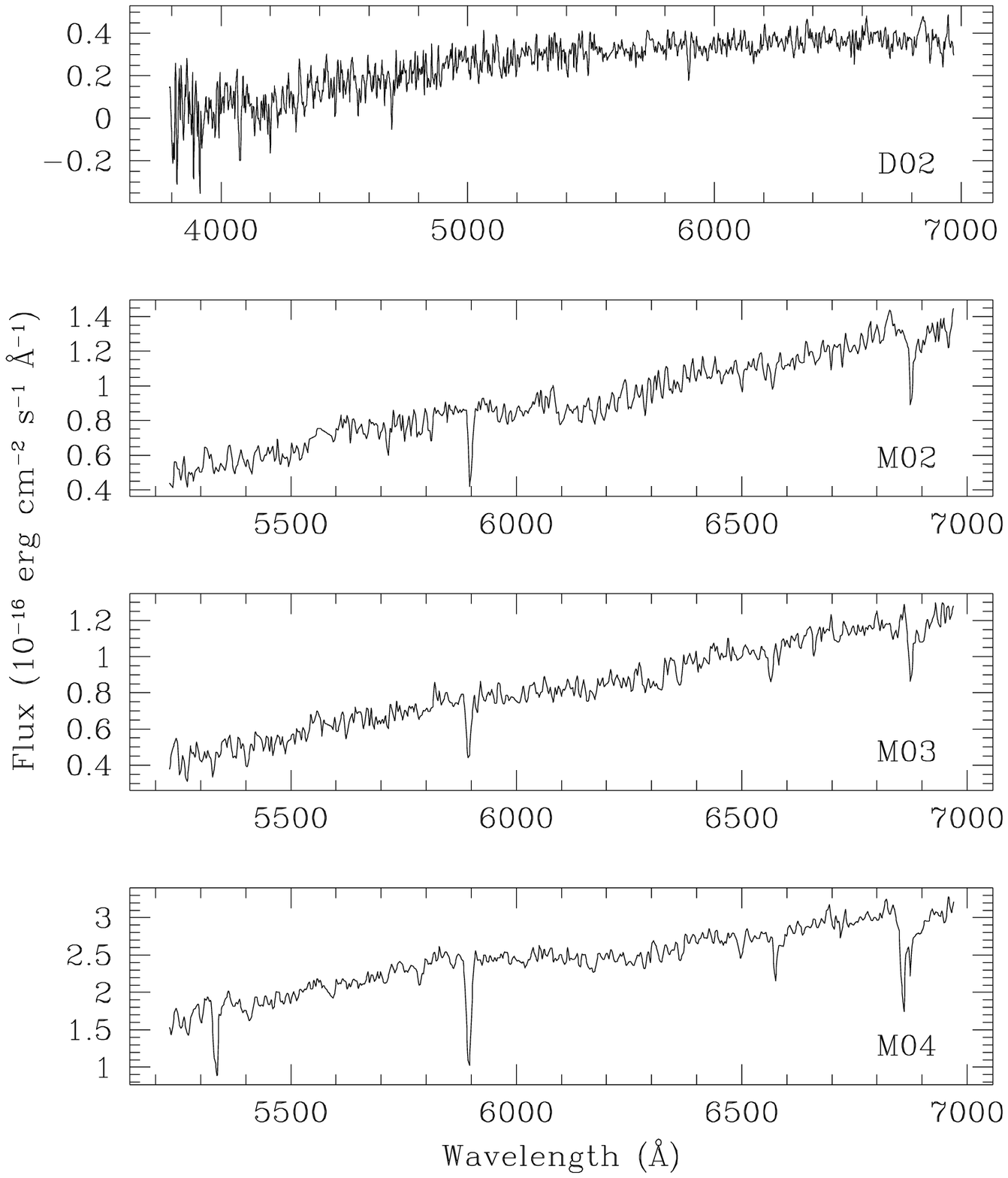}
\figcaption{Spectra of the microlensed sources DUO 95-BLG-2,
MACHO 95-BLG-2, MACHO 95-BLG-3, and MACHO 95-BLG-4.}
\end{figure}}

{\begin{figure}
\plotone{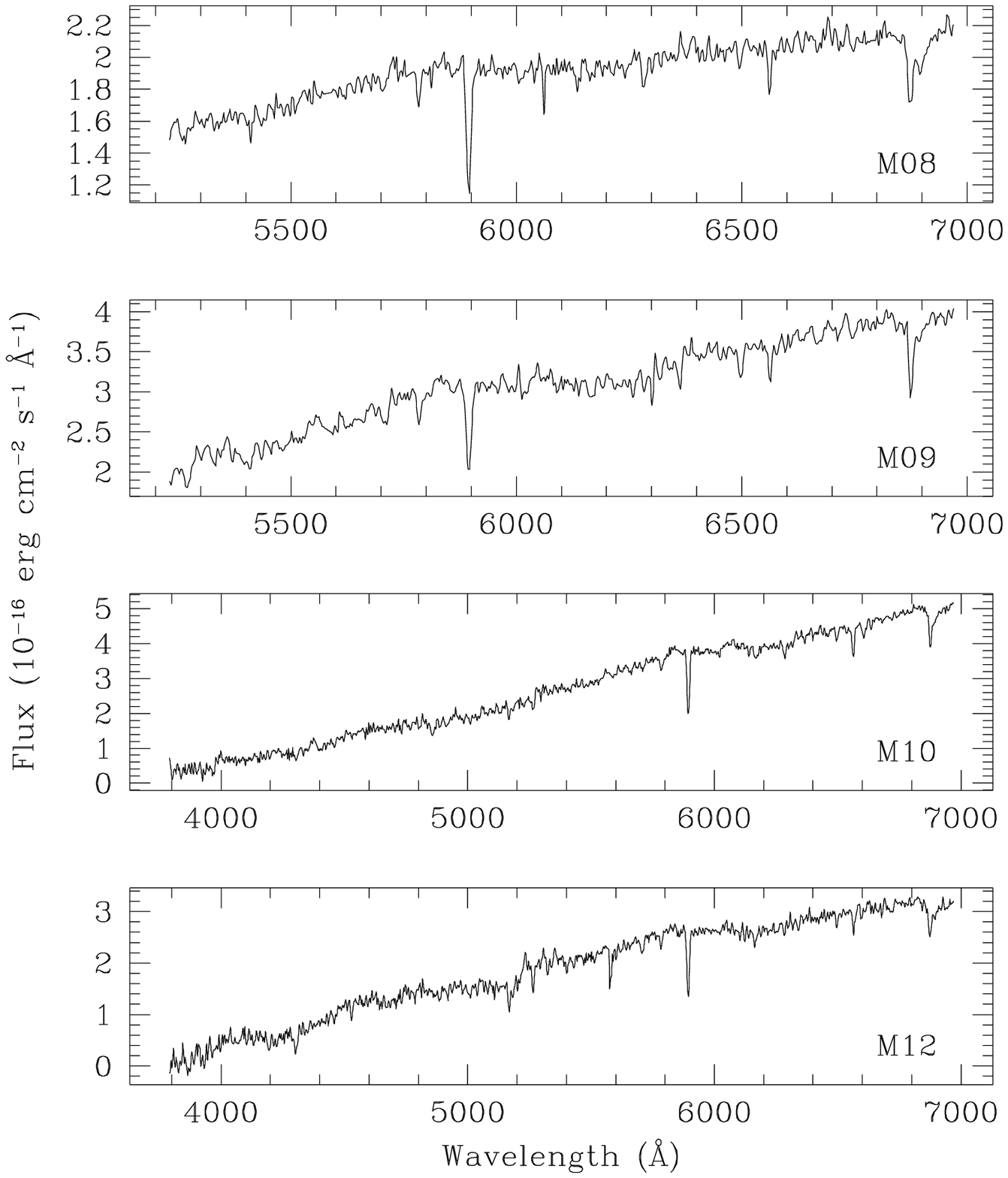}
\figcaption{Spectra of the microlensed sources MACHO 95-BLG-8,
MACHO 95-BLG-9, MACHO 95-BLG-10, and MACHO 95-BLG-12.}
\end{figure}}

{\begin{figure}
\plotone{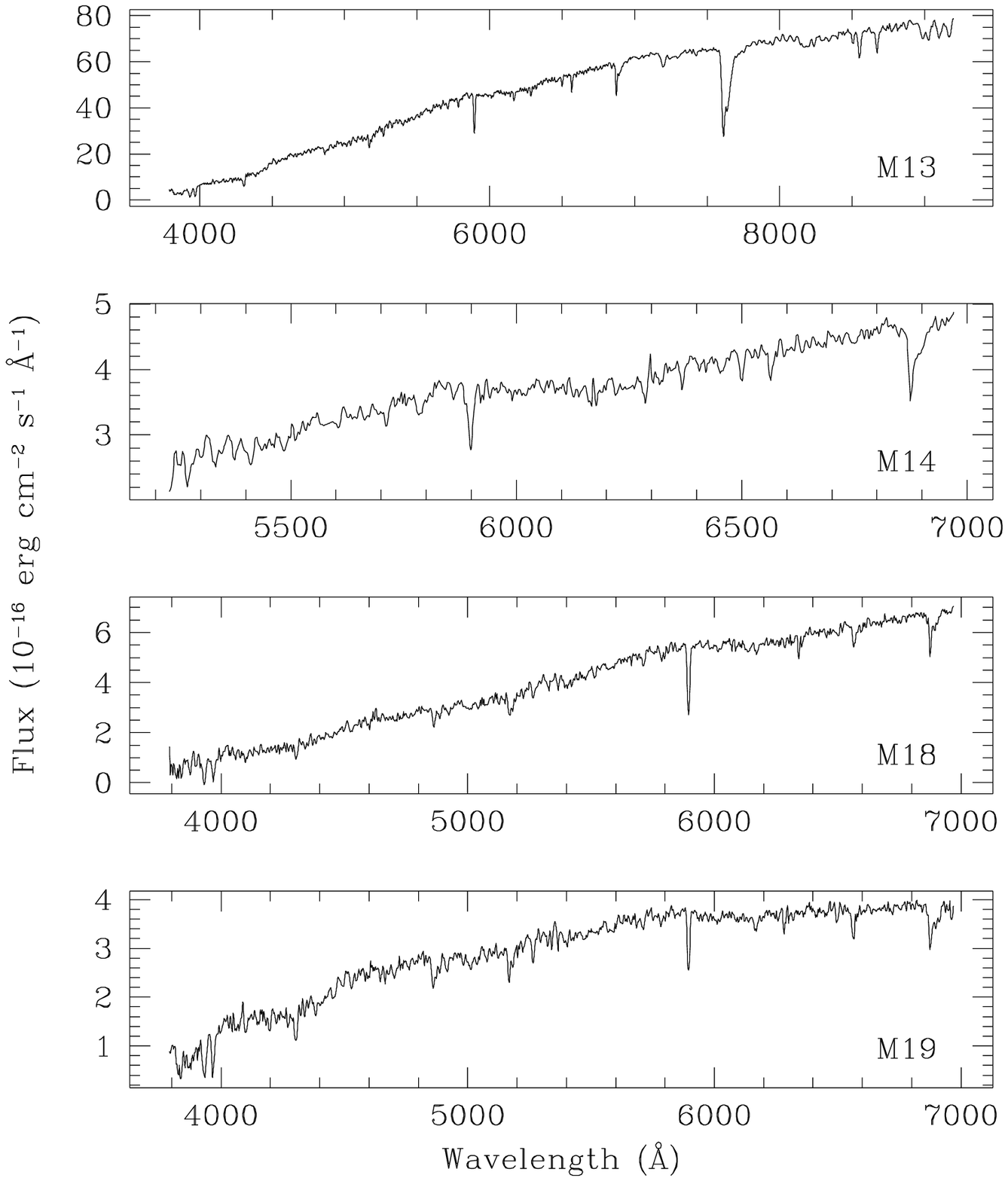}
\figcaption{Spectra of the microlensed sources MACHO 95-BLG-13,
MACHO 95-BLG-14, MACHO 95-BLG-18, and MACHO 95-BLG-19.}
\end{figure}}

{\begin{figure}
\plotone{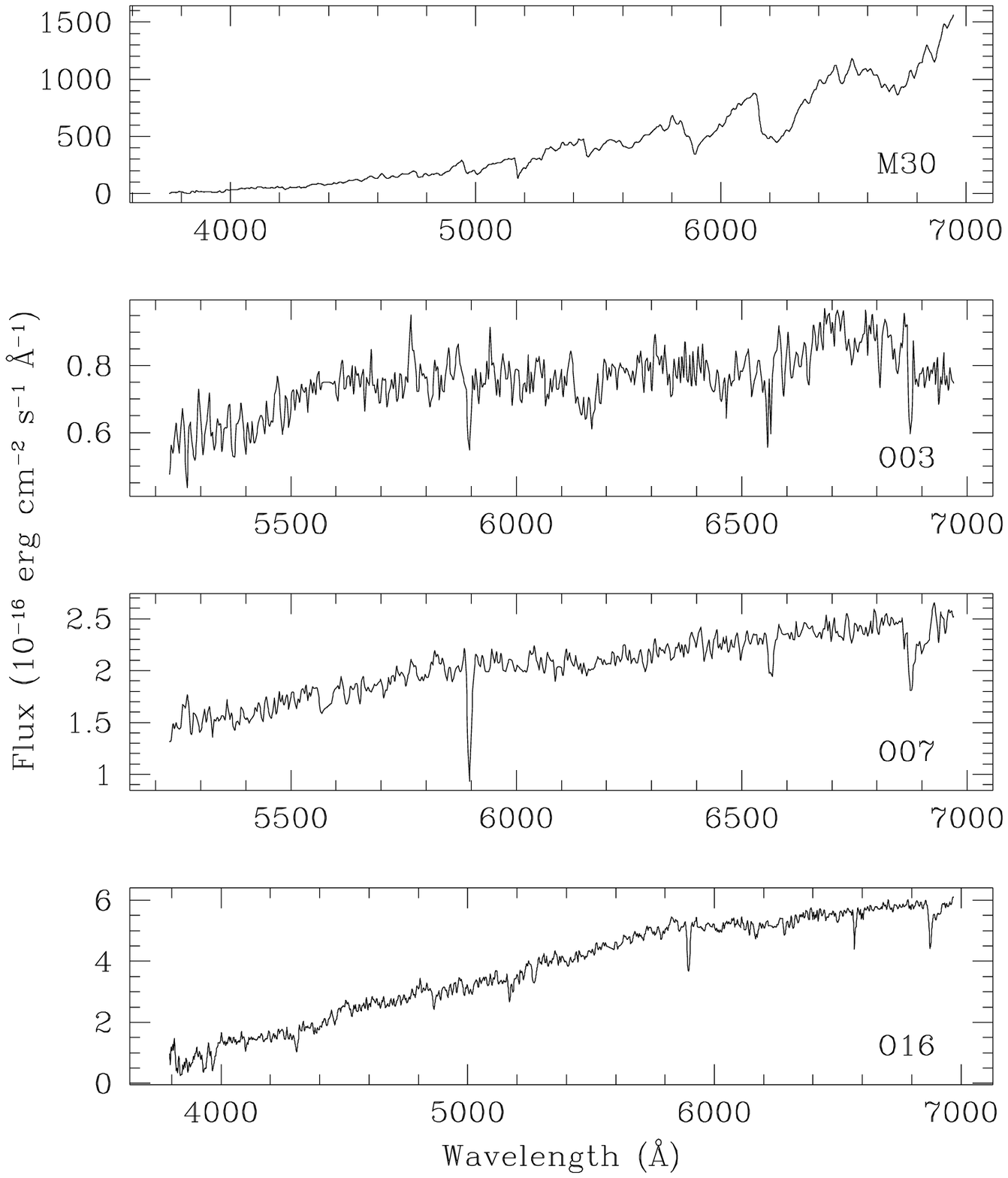}
\figcaption{Spectra of the microlensed sources MACHO 95-BLG-30,
OGLE 95-BLG-3, OGLE 95-BLG-7, and OGLE 95-BLG-16.}
\end{figure}}

Unfortunately, our wavelength coverage is 3500-7000 \AA, which is
different from that of the MACHO collaboration. Furthermore, the
spectrum by the MACHO group was obtained when the source was amplified.
So the blending fraction (i.e., fraction of light from a possible
blended object) may be different at the two epochs which may affect the
result. And, as explained earlier, the limitations in the theoretical
models make our spectral classifications uncertain if the spectral type
is about M2 or later. 

It would be interesting to compare our spectral classification with
other such estimates available in the literature. There is one such
source, MACHO 95-BLG-30, which has been studied in detail by the MACHO
collaboration \citep{alc97} who obtained spectra with a wavelength
coverage of 6230--9340 \AA. From this they estimated a spectral type of
M4III, which is close to and within uncertainties of our determination
of M2III.

As explained earlier, our velocity determinations are relative and the
absolute velocity can have large uncertainties because of the
uncertainty in the absolute velocity of the template star. However,
we are interested only on relative velocities in this study,
which are more accurate as explained before.

\subsection{Color-Magnitude Diagram Analysis}

In carrying out a comparative study of the properties of the microlensed
and non-microlensed sources, it is important to check the distributions
of both samples of the sources in the color-magnitude diagram (CMD)
since they can provide some insight into the sample being analyzed.

There have been several studies performed on Galactic bulge CMDs, such
as \citet{ter88} who was the first to use a CCD in this analysis. The
OGLE collaboration has since presented CMDs of 14 fields surveyed in
the direction of the Galactic bulge \citep{uda93}. Some of the features
common to these CMDs have been further studied, such as the
well-defined red clump branch \citep{sta94} and the distribution of
the disk stars \citep{pac94a}.

{\begin{figure}
\plotone{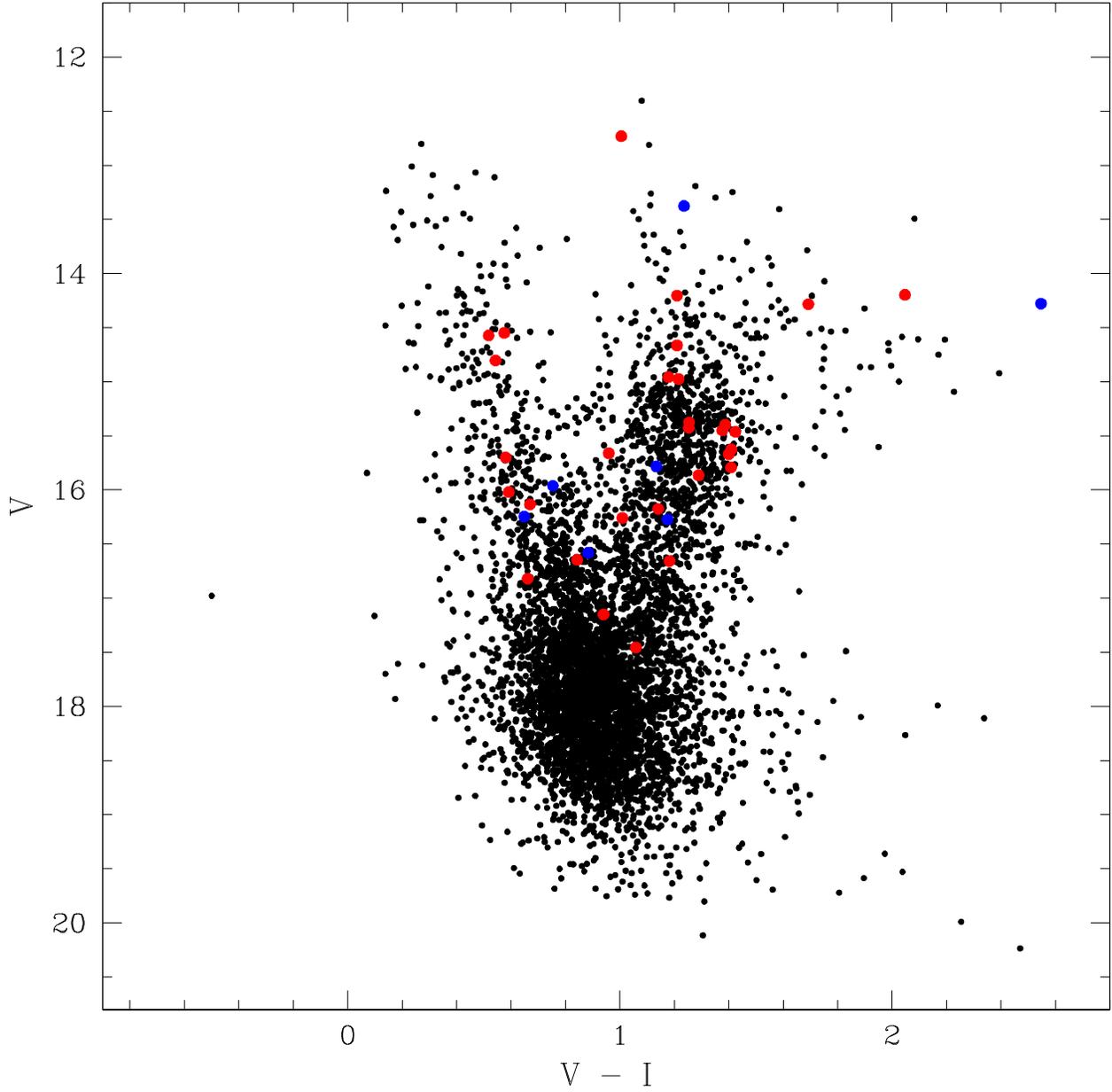}
\figcaption{Combined color-magnitude diagram for the MACHO bulge
fields 10, 12, 13, 17, 18, 19, and 30. The microlensed sources are shown
in blue and the non-microlensed sources are shown in red.}
\end{figure}}

During the 1995 observing season, photometric data from 7 of the 17
fields studied in this paper were obtained by the PLANET collaboration
\citep{alb98}. The CMDs for the individual fields have been combined
into a single CMD, as shown in Figure 10. The combined CMD contains
almost 22000 stars from the MACHO bulge fields 10, 12, 13, 17, 18, 19,
and 30. Since the distribution of stars in the individual CMDs was
almost identical, the colors and magnitudes were calibrated by adopting
the position for the bulge red clump giants as estimated by
\citet{pac98}, who found an average $(V-I)$ for the red clump region of
1.22 and an average $I$ magnitude of 14.34. The microlensed sources are
shown in blue in the CMD and the non-microlensed sources are shown in
red.

The combined CMD is in good agreement with the CMDs published by OGLE.
As expected, the CMD is dominated by bulge stars contained in a wide
main sequence turnoff point and the red giant branch. Also visible in
the diagram is a high concentration of stars in the blue part of the
CMD, suggested to be dominated by disk stars \citep{pac94a}. The
non-microlensed stars chosen for this study are of similar brightness
to the microlensed sources. The combined CMD shows that these stars lie
within the same sample as the microlensed sources and are generally
located in the recognizable main sequence or red giant branch. Hence,
the non-microlensed stars chosen for comparison in this study are fairly
typical of the population towards the Galactic bulge and are suitable
for use in this study.

It is of interest to compare the magnitudes and colors of the sources
as derived from the CMD with the spectral classifications derived from
the spectra. Shown in Table 7 are these results along with an estimate of
the absolute magnitude $M_V$. The value of $M_V$ derived here assumes
that the peak of the red clump is at $m_I = 14.34$ and $(V-I) = 1.22$, as
found by \citet{pac98}. This is equivalent to assuming that the star is
approximately in the middle of the bulge corresponding to a distance
modulus of 14.62. The error in $M_V$ is dominated by the intrinsic
dispersion of the bulge stars, which is $\sigma_V \sim 1.5$ magnitudes.
These results show that the luminosities measured from the CMD are
roughly consistent with the classifications derived from the spectra and
that giant stars have been preferentially selected since they are most
likely to be in the Galactic bulge. Note that the magnitudes and colors
shown in Table 7 are not the calibrated values but they are the expected
dereddened magnitudes and colors if the sources were at the middle of the
Galactic bulge. This also shows that the non-microlensed stars chosen
here are also within the bulge and hence they form a good sample for our
comparative study.  

\subsection{Extinctions of Microlensed vs. Non-microlensed Sources}

Although the internal extinction within Baade's window is thought to be
small, there are large uncertainties. As discussed earlier, the
microlensed sources will show an extinction offset relative to the
unlensed stars if the extinction within the bulge is non-negligible. 
We note that almost all the sources (the microlensed as well as the
comparison stars) are expected to be within the bulge, and that the
sources lie within a fairly restricted region of the bulge
($0.55\degr<l<3.98\degr$ and $-4.92\degr<b<-2.68\degr$). Hence the {\it
foreground} extinction caused by the Galactic disk (i.e., excluding the
extinction within the bulge) is expected to be the same for all the
sources. Thus, any extinction offset between the non-microlensed and
microlensed samples would indicate that extinction within the bulge is
non-negligible. As explained in paper I, this extinction offset can be
used to estimate the fraction of bulge-bulge lensing.

{\begin{figure}
\plotone{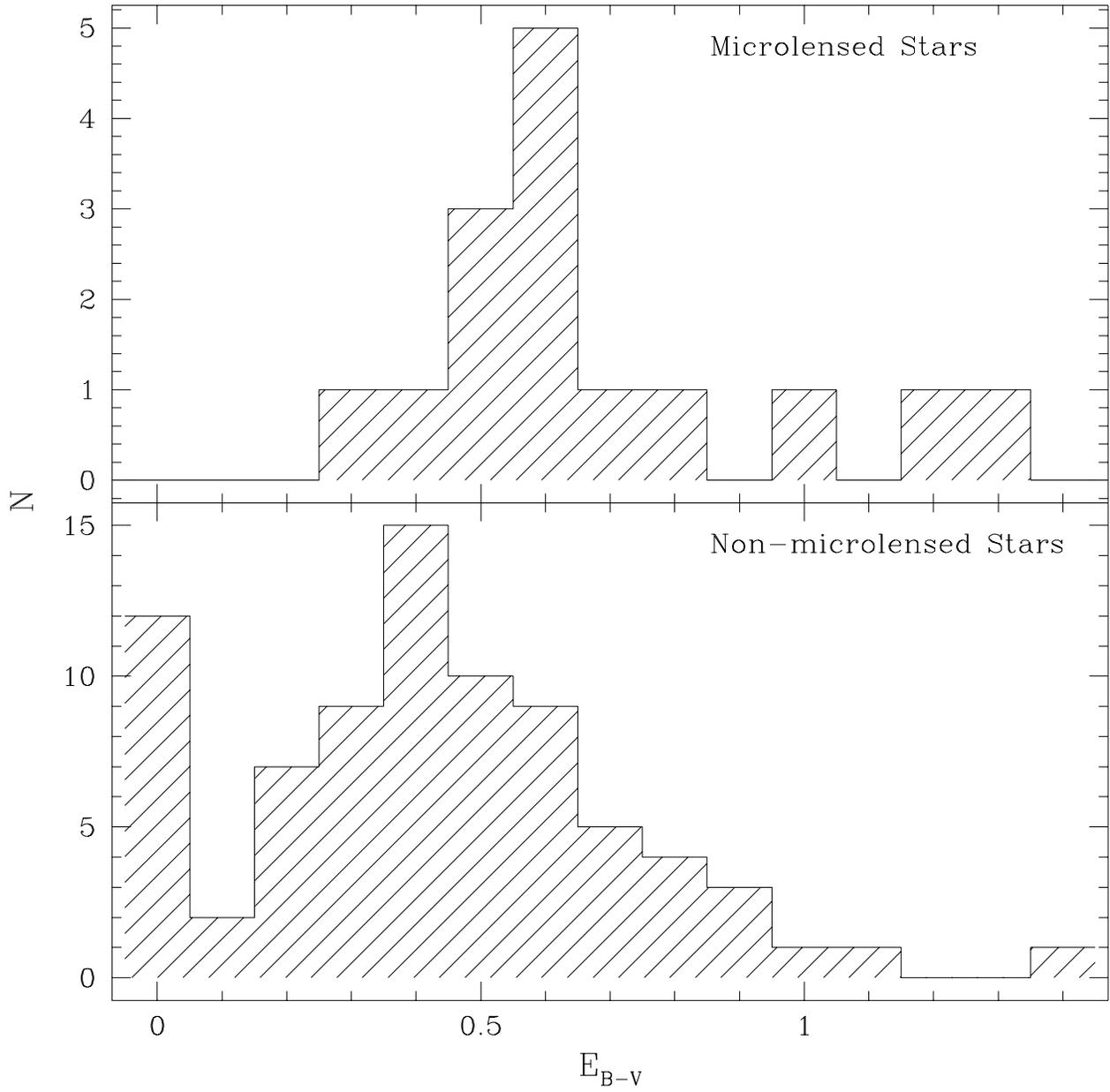}
\figcaption{Histogram of extinction values for microlensed and
non-microlensed stars.}
\end{figure}}

Shown in Figure 11 is a histogram of the extinction for microlensed and
non-microlensed stars. There is a surprisingly large number of stars
with little or no extinction amongst the non-microlensed stars which
is most likely due to disk stars of low mass. This could be explained
in part by the findings of \citet{pac94a} that indicate that there is
an excess of disk stars by a factor of $\sim 2$ between us and a
distance of 2.5 kpc towards the Galactic bulge, and a rapid drop by a
factor of $\sim 10$ beyond that distance. The average extinction for
the microlensed sources is $E_{B-V} = 0.68$ and is $E_{B-V} = 0.43$ for
the non-microlensed stars. As expected, the distributions peak at these
average values. The offset between the two mean values of $\Delta
E_{B-V} = 0.25$ is equivalent to a magnitude offset of $\approx 0.80$
in $V$.

To investigate the significance of the offset between the two mean
values, a t-test was performed on the histogram data. A value of $t =
3.07$ was obtained for 93 degrees of freedom which results in a
probability of $p = 0.01$. In other words, the difference in the mean
values of the two distributions is significant at the 99\% confidence
level. To test how much weight is held by the unlensed stars with zero
extinction, the t-test was performed again after removing these stars
from the sample. This reduced the values to $t = 2.42$ for 81 degrees
of freedom which results in a probability of $\approx 0.02$, or
significance at the 98\% confidence level.

This test assumes a normal distribution for the data sets which is
difficult to determine given the relatively small number of
microlensed sources included in this sample. These results appear
to agree with the previous discussions regarding the extinction
bias of microlensed sources and a clear trend is seen in the
presented histogram. 

As explained in paper I, from the extinction distribution for
microlensed sources it is possible to make an estimate of the fraction
of bulge-bulge lensing. Using the formalism of paper I, a simple
estimate of the fraction of bulge-bulge lensing is found to be
$\sim 65\%$. This is consistent with earlier predictions based on the
presence of the bar \citep{pac94b,zha95}.

We now turn to another possible effect: the time scales of the
microlensing events versus the extinction. For self-lensing within the
bulge, the Einstein ring size increases if the distance between the
lens and the source is larger. Since the microlensed sources are
preferentially located at the far side of the bulge, the
characteristic time scale should be longer for events exhibiting
larger extinction if the internal extinction is important. However,
the time scale of an event is also a function of the velocities of the
lens and the source and so the time scale may depend upon the Galactic
kinematics. If the velocity dispersion is the dominant component and
is similar in different regions of the bulge then the time scale should
be larger for a higher value of extinction. On the other hand, if
rotation is the dominant component then the time scale may show a
behavior which is only a small function of the extinction value.

{\begin{figure}
\plotone{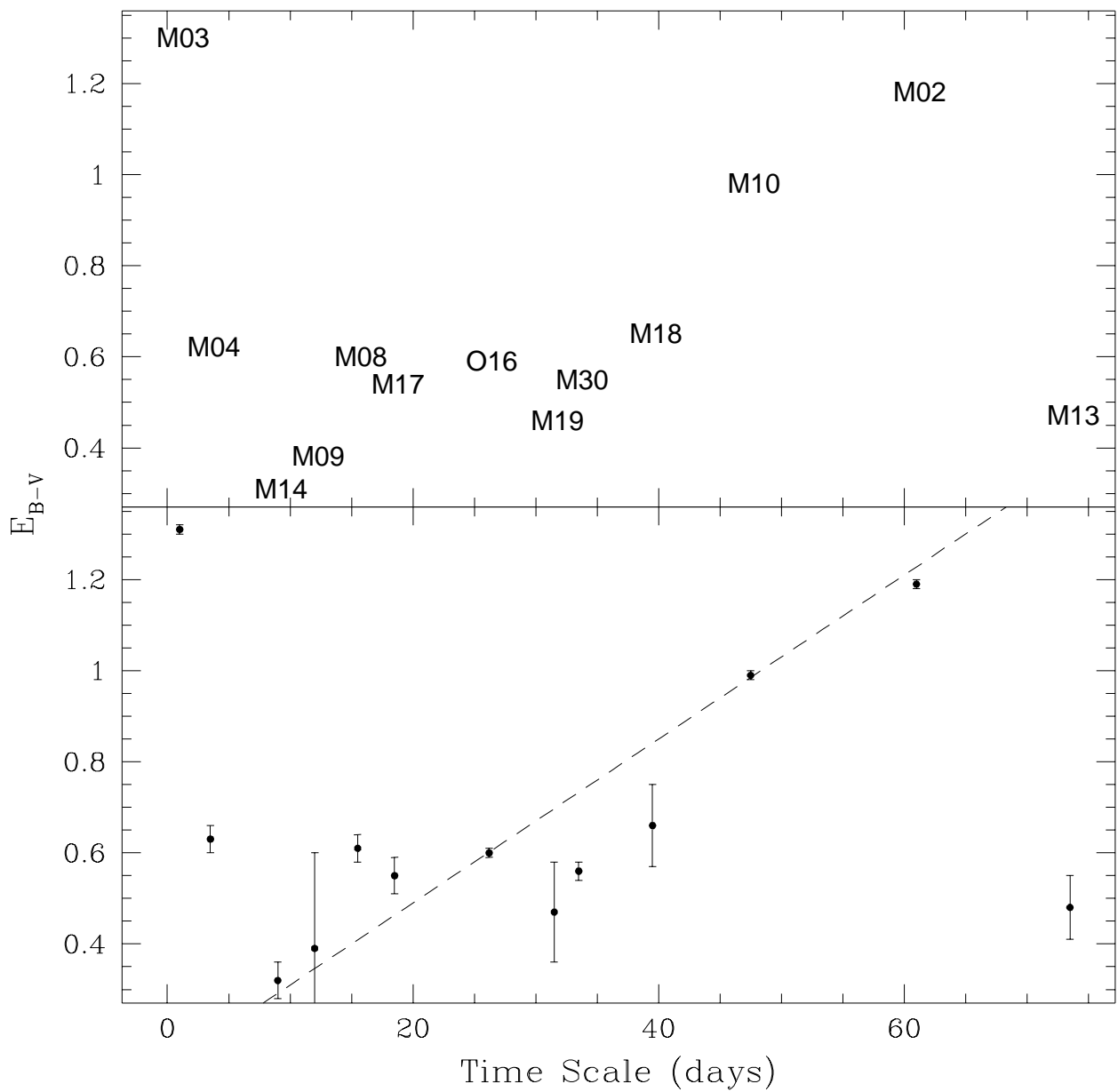}
\figcaption[f12.eps]{Plot of the extinction of microlensed sources as a
function of their characteristic time scales. Events exhibiting binary
behavior are not included in this plot. {\bf Top:} An abbreviated
version of the event names is used to indicate the source positions.
{\bf Bottom:} A line of best fit to the points excluding events
MACHO 95-BLG-3 and MACHO 95-BLG-13.}
\end{figure}}

Shown in Figure 12 is a plot of the extinction of the microlensed
sources as a function of their characteristic time scales. The top
frame uses the abbreviated names of the events to show their positions
on the plot and the bottom frame shows the corresponding data points
with error bars. We should note that MACHO 95-BLG-13 is a very bright
source and it has relatively low extinction. Therefore, it is almost
certainly a disk star and hence should not be included in this
analysis. MACHO 95-BLG-3 has an extremely short time scale which could
mean that the source and the lens are very close to each other (both at
the far side of the bulge), rather than a very small mass of the lens
or a very high relative velocity. As indicated in paper I, this would
make the event quite unusual and not typical of Galactic microlensing.

Rejecting these two anomalous points for the reasons expressed above, a
line was fitted to the data using linear regression. This fit is shown
in the bottom panel of Figure 12. The trend in this data shows that the
velocity dispersion component of the Galactic kinematics is strong
enough such that there is a correlation between the extinction and the
characteristic time scale of the event. Linear regression was used to
obtain a linear fit which produces the following equation for this
trend 
\begin{equation}
E_{B-V} = 0.018 \, t_E + 0.13
\end{equation}
Of course, contamination due to disk lensing will cause greater scatter
in this result. This clear trend seems to further confirm the earlier 
result that the microlensed sources suffer from larger extinction and 
that they are predominantly at the far side of the bulge. We emphasize
however, that the uncertainty in individual extinction measurements can
be large. Furthermore, although there seems to be a clear linear trend,
the trend is dominated by just two points, namely MACHO 95-BLG-02 and
MACHO 95-BLG-10. Further observations will help in confirming this result.

\subsection{Kinematics of Microlensed Sources}

The microlensed sources provide a unique opportunity to investigate the
kinematic properties in this region. The kinematic properties of our
Galaxy have been studied and used with microlensing to construct
consistent models of the Galaxy \citep{mer98}. It was suggested by
\citet{wal97} that one could use the kinematic properties of
microlensed sources to distinguish between a lens population in the
disk and a lens population in the bulge. It was also found that the
radial velocity distributions between the two populations should not be
substantially different for an axisymmetric bulge model but may exhibit
a relative shift if the bulge is non-axisymmetric (barred), depending
upon the kinematics of the bar. As we found in the previous section,
the distribution of transverse velocities of these two populations
should be substantially different.

{\begin{figure}
\plotone{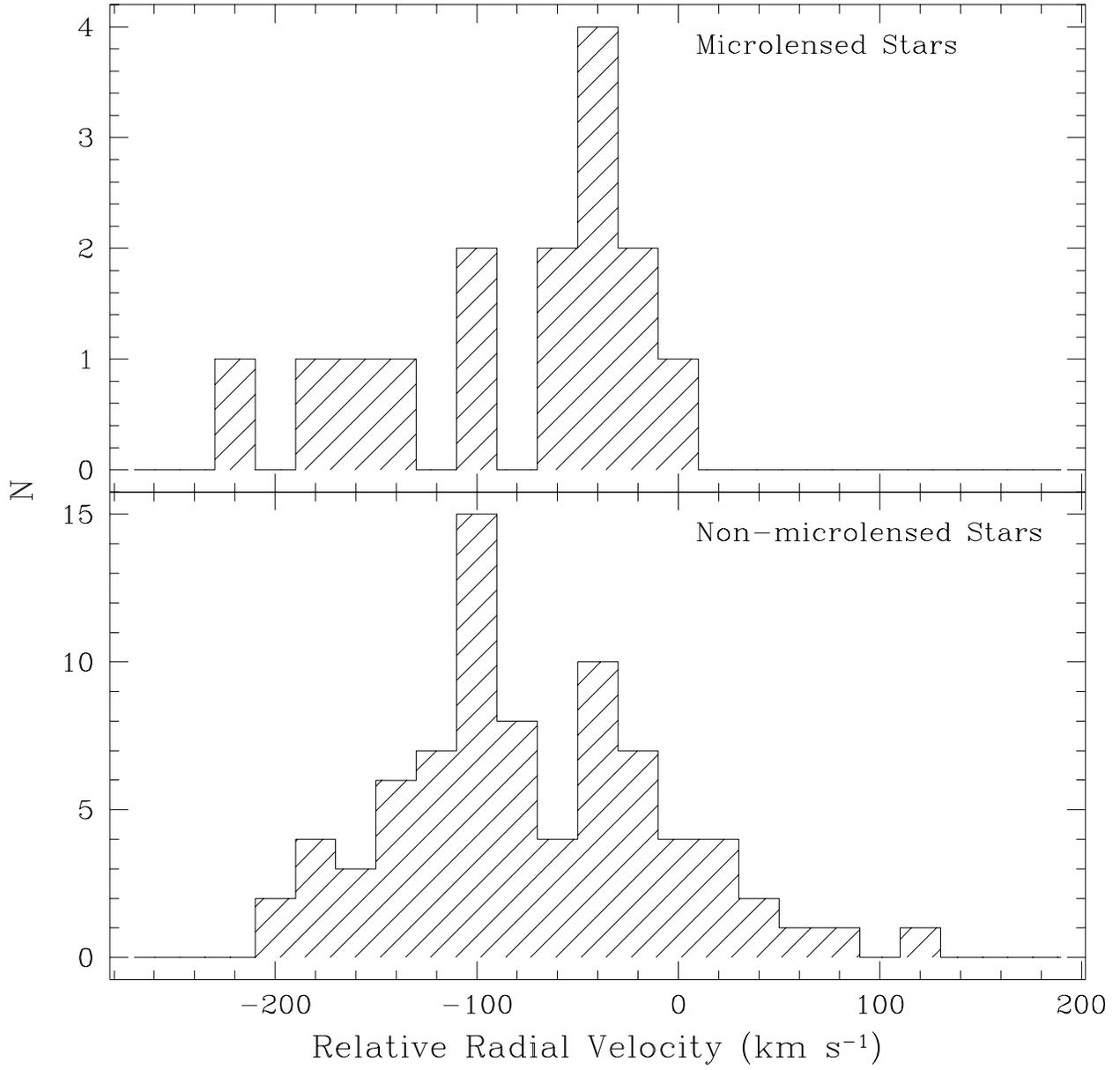}
\figcaption[f13.eps]{Histogram of relative radial velocities for
microlensed and non-microlensed stars.}
\end{figure}}

Shown in Figure 13 is a histogram of the relative radial velocities for
microlensed and non-microlensed stars. The average relative radial
velocity for the microlensed sources is $v_r = -81.2$ and is $v_r =
-71.5$ for the non-microlensed stars. There appears to be a slight
shift between the peak values of these two distributions but this is
well within the velocity dispersions of the sources, and given the
small number of samples and the low-resolution of the spectra, this 
is not statistically significant.

To investigate the significance between the two mean values, a t-test
was performed on the histogram data. A value of $t = 0.52$ was obtained
for 92 degrees of freedom which results in a very high probability that
the difference in the mean values of the two distributions is not
statistically significant.

This lack of correlation could be due to either of the following:
(i) there is no intrinsic correlation, or (ii) the uncertainties in
the velocity measurements is larger than estimated here, or
(iii) the resolution is insufficient to see any correlation.
Clearly, more spectroscopic observations with better S/N at a higher 
resolution will help in finding the exact cause.

We note that there was a significant difference between the extinction
distributions for the microlensed and non-microlensed stars. Since
there is no such difference between the radial velocity distributions,
it is not expected that there will be a correlation between the
extinction and radial velocity of the observed sources.

{\begin{figure}
\plotone{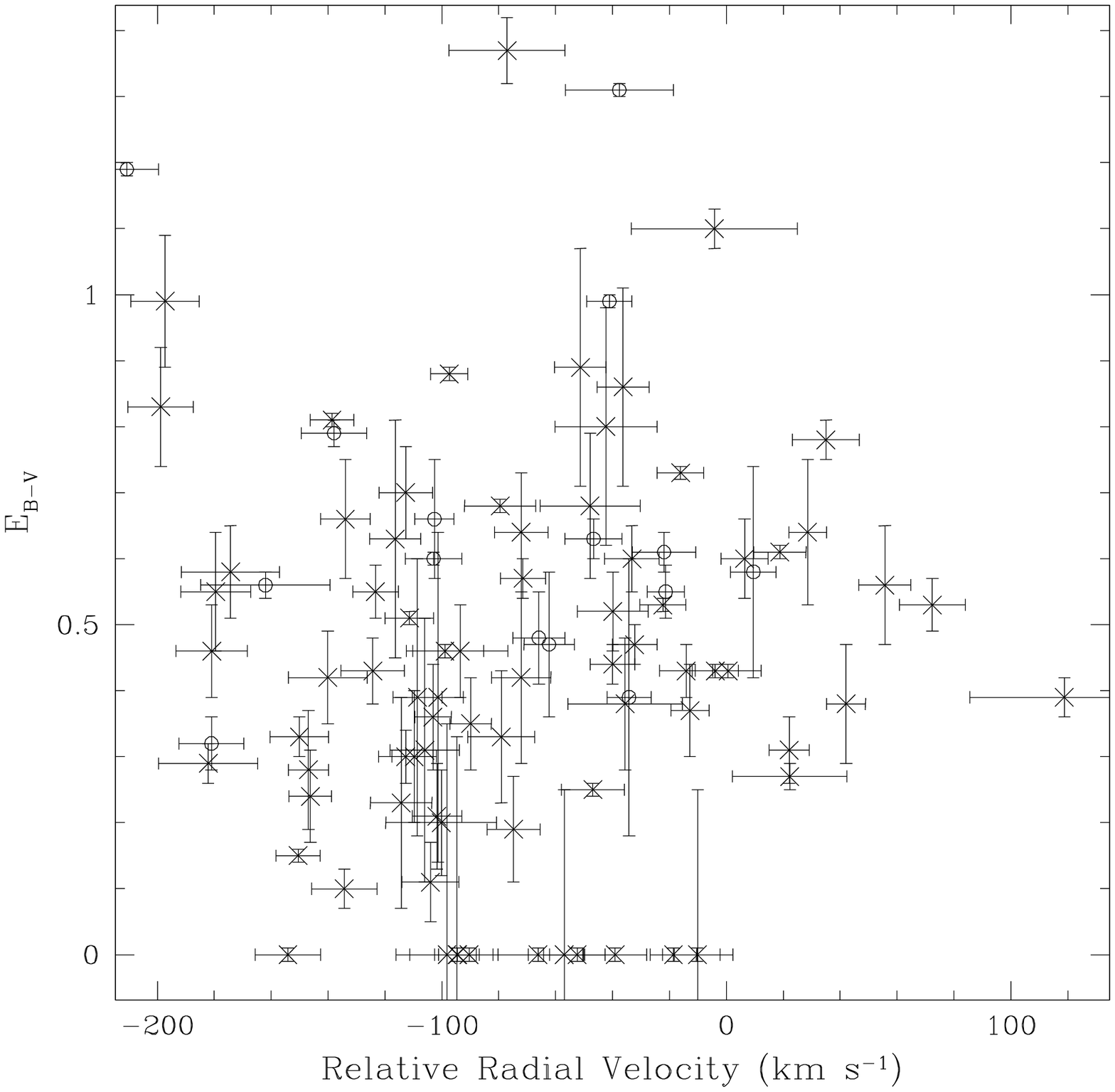}
\figcaption[f14.eps]{Plot of the extinction of all observed sources as a
function of their relative radial velocities. The microlensed sources
are shown as open circles and the non-microlensed sources are shown as
crosses.}
\end{figure}}

Shown in Figure 14 is a plot of the extinction of the microlensed and
non-microlensed stars as a function of their relative radial velocities.
This figure excludes only one star, namely star 3 from the MACHO
95-BLG-30 field, for which there were not enough spectral lines in the
star's B150 spectrum for an accurate estimate of the radial velocity.
(For this source, the routine produced a relative radial velocity
measurement of $v_r = -399.8 \pm 14.1$, but this may be because of a
misidentification of some spectral lines).

The microlensed sources are shown as circles and the non-microlensed
sources are shown as crosses. As expected, there is an extinction
shift between the microlensed population and the non-microlensed
population but there is no apparent correlation between the
extinction and the radial velocity. These results appear to agree
with the postulates made by \citet{wal97}. These results are also
expected since the total extinction varies depending upon the line of
sight.

\section{CONCLUSIONS}

We have presented spectra of 17 microlensed sources taken with the
EFOSC at the ESO 3.6m telescope. These spectra were used to derive 
the spectral type, extinction, and the radial velocities of these
stars. Spectra were also taken of many non-microlensed sources in the
same fields. The same analysis was done for the non-microlensed
sources, and their spectral types, extinctions, and radial velocities
were determined. This was carried out by developing MIDAS scripts to
provide estimates of the extinction, spectral type, and radial
velocity for each individual spectrum. A large library of Kurucz model
spectra was constructed to model the spectra, and the radial velocities
were measured relative to a bright star using the cross-correlation
technique. These results are used for a comparative study of the
physical properties of the microlensed and non-microlensed stars.

A comparison of the extinction distributions for microlensed and
non-microlensed stars have been carried out through a statistical
analysis. The average extinction for the microlensed sources is
$E_{B-V} = 0.68$ and is $E_{B-V} = 0.43$ for the non-microlensed stars.
The offset between the two mean values of $\delta E_{B-V} = 0.25$
corresponds to a magnitude offset of $\delta A_v \approx 0.80$. A t-test
performed on these distributions showed that the difference in the mean
values of the two distributions is significant at the 99\% confidence
level.

A plot of the extinction of the microlensed sources as a function of
their characteristic time scales shows that the sources with larger
extinction, in general, correspond to larger time scales. This is
consistent with the expectation that the sources with larger extinction
lie farther along the line of sight.

The histogram presenting the relative radial velocities for microlensed
and non-microlensed stars shows that the difference between the two
distributions is not statistically significant. The sample needs to be
increased to about 100 sources to detect any possible offset. Thus, more
spectra of microlensed sources will be very useful in modeling the
kinematics of the Galactic bulge. It is expected, however, that the
transverse velocities of these two populations would be different. Hence
it would be of great interest to determine the transverse velocities of
the microlensed sources as these would, when combined with the radial
velocities, contribute significantly to the knowledge of the kinematics
of the far side of the Galactic bulge. It should be possible to make
such measurements with the Hubble Space Telescope or with future space
telescopes, such as NGST.

An estimate of the fraction of bulge-bulge lensing was made from the
extinction distribution for microlensed sources. This simple method
provides a rough estimate of the fraction of bulge-bulge lensing and is
found to be $\sim 65\%$. This value is similar to the results obtained
by previous investigations \citep{pac94b,zha95,kir94}.

\acknowledgements

The authors would like to thank Dr. Mike Bessell for his comments
regarding the spectral analysis and the PLANET collaboration for
making available the data from the 1995 PLANET observing season.
This research was financially supported by a DDRF grant of the
Space Telescope Science Institute.

\clearpage

\begin{deluxetable}{lrrr}
\tablecaption{Grisms used, their characteristics, and the rms deviations
from the dispersion relation.}
\tablewidth{0pt}
\tablehead{
\colhead{Grism} & \colhead{Wavelength Range (\AA)} &
\colhead{Dispersion (\AA/pixel)} & \colhead{rms (pixels)}
}
\startdata
B150 & 3780--5510 & 3.3 & 0.18\\
B300 & 3740--6950 & 6.3 & 0.31\\
O150 & 5220--6980 & 3.4 & 0.19\\
R300 & 5890--9920 & 7.9 & 0.20\\
\enddata
\end{deluxetable}

\clearpage

\begin{deluxetable}{lccccr}
\tablecaption{Microlensing event information provided by the MACHO alerts
and the OGLE and DUO publications. Information includes the coordinates
of the sources, the baseline $V$ and $R$ magnitudes, and the
characteristic time scales of the events.}
\tablewidth{0pt}
\tablehead{
\colhead{Event} & \colhead{R.A. (J2000)} & \colhead{DEC. (J2000)} &
\colhead{V} & \colhead{R} & \colhead{$t_E$ (days)}
}
\startdata
DUO-1995-BLL-2 & 18:10:17.2 & -27:28:49 & - & 18.6 & binary\\
MACHO 95-BLG-2 & 18:08:25.2 & -27:58:38 & 19.0 & 17.9 & 61.0\\
MACHO 95-BLG-3 & 18:02:37.5 & -29:39:36 & 18.7 & 17.6 & 1.0\\
MACHO 95-BLG-4 & 18:00:03.4 & -29:11:04 & 18.1 & 17.1 & 3.5\\
MACHO 95-BLG-8 & 18:16:46.0 & -26:11:43 & 17.1 & 16.3 & 15.5\\
MACHO 95-BLG-9 & 18:06:32.3 & -30:55:55 & 17.2 & 16.2 & 12.0\\
MACHO 95-BLG-10 & 17:58:16.0 & -29:32:11 & 18.9 & 18.0 & 47.5\\
MACHO 95-BLG-12 & 18:06:04.8 & -29:52:38 & 18.6 & 17.7 & binary\\
MACHO 95-BLG-13 & 18:08:47.0 & -27:40:47 & 16.6 & 15.6 & 73.5\\
MACHO 95-BLG-14 & 18:01:26.3 & -28:31:14 & 17.4 & 16.5 & 9.0\\
MACHO 95-BLG-17 & 18:03:01.1 & -28:21:09 & 18.8 & 18.0 & 18.5\\
MACHO 95-BLG-18 & 18:07:20.6 & -28:36:51 & 18.7 & 17.8 & 39.5\\
MACHO 95-BLG-19 & 18:11:32.5 & -27:45:27 & 18.6 & 17.9 & 31.5\\
MACHO 95-BLG-30 & 18:07:04.3 & -27:22:06 & 16.1 & 14.7 & 33.5\\
OGLE 95-BLG-3 & 18:04:43.5 & -30:14:11 & 17.7 & - & 11.5\\
OGLE 95-BLG-7 & 18:03:35.8 & -29:47:06 & 19.3 & - & binary\\
OGLE 95-BLG-16 & 18:02:07.6 & -30:01:12 & 20.0 & - & 26.2\\
\enddata
\end{deluxetable}

\clearpage

\begin{deluxetable}{lccc}
\tablecaption{Summary of microlensing events observed. Information includes
the date of observation (HJD - 2449000), the grisms used, and the
respective exposure times.}
\tablewidth{0pt}
\tablehead{
\colhead{Event} & \colhead{Obs. Date} & \colhead{Grisms Used} &
\colhead{Exposure (seconds)}
}
\startdata
DUO 95-BUL-2 & 899.64 & B150,O150 & 1800,2220\\
MACHO 95-BLG-2 & 899.89 & O150 & 2280\\
MACHO 95-BLG-3 & 899.86 & O150 & 1200\\
MACHO 95-BLG-4 & 899.84 & O150 & 1800\\
MACHO 95-BLG-8 & 899.71 & O150 & 2700\\
MACHO 95-BLG-9 & 899.78 & O150 & 1800\\
MACHO 95-BLG-10 & 897.79 & B150,O150 & 1800,1800\\
MACHO 95-BLG-12 & 897.74 & B150,O150 & 1500,1200\\
MACHO 95-BLG-13 & 897.70 & B150,O150,R300 & 900,600,120\\
MACHO 95-BLG-14 & 899.81 & O150 & 1800\\
MACHO 95-BLG-17 & 898.69 & B150,O150 & 1200,1800\\
MACHO 95-BLG-18 & 898.73 & B150,O150 & 1800,1800\\
MACHO 95-BLG-19 & 897.85 & B150,O150 & 1800,1800\\
MACHO 95-BLG-30 & 1251.75 & B150,B300 & 1200,900\\
OGLE 95-BLG-3 & 899.75 & O150 & 1800\\
OGLE 95-BLG-7 & 898.84 & O150 & 2700\\
OGLE 95-BLG-16 & 898.79 & B150,O150 & 1800,1800\\
\enddata
\end{deluxetable}

\clearpage

\begin{deluxetable}{lccccc}
\tablecaption{Coefficients for polynomial fits to $\log g$ values.}
\tablewidth{0pt}
\tablehead{
\colhead{Class} & \colhead{a} & \colhead{b} & \colhead{c} &
\colhead{d} & \colhead{e}
}
\startdata
I & $1.71 \times 10^{-3}$ & -0.05 & 0.37 & -1.50 & 4.05\\
III & $5.75 \times 10^{-3}$ & -0.06 & 0.04 & 0.44 & 2.91\\
V & $9.30 \times 10^{-3}$ & -0.13 & 0.56 & -0.80 & 4.30\\
\enddata
\end{deluxetable}

\clearpage

\begin{deluxetable}{lccc}
\tablecaption{Classification, extinction, and relative radial velocity
results for each of the observed microlensed sources.}
\tablewidth{0pt}
\tablehead{
\colhead{Spectrum} & \colhead{Class} & \colhead{$E_{B-V}$} &
\colhead{$v_r \, \mathrm{(km \, s^{-1})}$}\\
}
\startdata
D02 & K2III & $0.22 \pm 0.09$ & $-97.3 \pm 16.6$\\
M02 & G2III & $1.19 \pm 0.01$ & $-210.7 \pm 11.2$\\
M03 & G0III & $1.31 \pm 0.01$ & $-37.6 \pm 19.0$\\
M04 & K0III & $0.63 \pm 0.03$ & $-46.7 \pm 10.0$\\
M08 & G0III & $0.61 \pm 0.03$ & $-21.9 \pm 11.2$\\
M09 & K4III & $0.39 \pm 0.21$ & $-34.2 \pm 7.8$\\
M10 & G2III & $0.99 \pm 0.01$ & $-41.1 \pm 7.9$\\
M12 & K0III & $0.58 \pm 0.16$ & $9.5 \pm 8.0$\\
M13 & K1I & $0.48 \pm 0.07$ & $-65.9 \pm 9.1$\\
M14 & K4III & $0.32 \pm 0.06$ & $-181.0 \pm 11.4$\\
M17 & G5III & $0.55 \pm 0.04$ & $-21.3 \pm 6.6$\\
M18 & K0III & $0.66 \pm 0.09$ & $-102.6 \pm 6.9$\\
M19 & G2III & $0.47 \pm 0.11$ & $-62.3 \pm 8.8$\\
M30 & M2III & $0.56 \pm 0.02$ & $-162.0 \pm 22.7$\\
O03 & K2I & $0.08 \pm 0.01$ & $-63.1 \pm 11.3$\\
O07 & G0III & $0.79 \pm 0.02$ & $-137.9 \pm 11.5$\\
O16 & G5III & $0.60 \pm 0.01$ & $-102.9 \pm 10.0$\\
\enddata
\end{deluxetable}

\clearpage

\begin{deluxetable}{lccc}
\tablecaption{Classification, extinction, and relative radial velocity
results for the MACHO-1995-BLG-17 field.}
\tablewidth{0pt}
\tablehead{
\colhead{Spectrum} & \colhead{Class} & \colhead{$E_{B-V}$} &
\colhead{$v_r \, \mathrm{(km \, s^{-1})}$}\\
}
\startdata
Source & G5III & $0.55 \pm 0.04$ & $-21.3 \pm 6.6$\\
Star 1 & K3III & $0.42 \pm 0.13$ & $-72.1 \pm 10.4$\\
Star 2 & K5III & $0.33 \pm 0.03$ & $-150.1 \pm 10.3$\\
Star 3 & K2III & $0.64 \pm 0.09$ & $-72.1 \pm 9.4$\\
Star 4 & G2III & $0.29 \pm 0.03$ & $-182.1 \pm 17.4$\\
Star 5 & K2III & $0.46 \pm 0.07$ & $-180.9 \pm 12.5$\\
\enddata
\end{deluxetable}

\clearpage

\begin{deluxetable}{lcccc}
\tablecaption{Classifications derived from spectra, dereddened apparent
magnitudes and colors, and an estimate of the absolute magnitude for
each of the microlensed and non-microlensed sources of interest
included in the CMD.}
\tablewidth{0pt}
\tablehead{
\colhead{Star} & \colhead{Classification} & \colhead{$m_V$} &
\colhead{$V-I$} & \colhead{$M_V$}
}
\startdata
MB95010 & G2III & 16.25 & 0.65 & 1.63\\
Star 1 & G0III & 16.13 & 0.67 & 1.51\\
Star 2 & K5III & 14.28 & 1.69 & -0.34\\
Star 4 & K5III & 15.67 & 1.40 & 1.05\\
Star 5 & K3III & 15.63 & 1.41 & 1.01\\
MB95012 & K0III & 16.27 & 1.18 & 1.65\\
Star 1 & G2III & 17.15 & 0.94 & 2.52\\
Star 2 & K2III & 15.86 & 1.29 & 1.24\\
Star 3 & K3III & 15.45 & 1.38 & 0.83\\
Star 4 & G5V & 16.82 & 0.66 & 2.20\\
Star 5 & G2III & 16.02 & 0.59 & 1.40\\
MB95013 & K1I & 13.38 & 1.24 & -1.24\\
Star 1 & K5III & 15.40 & 1.39 & 1.78\\
Star 2 & G2III & 14.96 & 1.18 & 0.34\\
Star 3 & K5III & 15.38 & 1.25 & 0.76\\
Star 4 & K5III & 15.46 & 1.42 & 0.84\\
Star 5 & K3III & 15.43 & 1.25 & 0.81\\
MB95017 & G5III & 16.58 & 0.88 & 1.94\\
Star 1 & G5III & 14.57 & 0.52 & -0.05\\
Star 3 & G5III & 15.70 & 0.58 & 1.08\\
MB95018 & K0III & 15.78 & 1.13 & 1.16\\
Star 1 & K4III & 14.66 & 1.21 & 0.04\\
Star 2 & K2III & 14.98 & 1.22 & 0.36\\
Star 3 & K4III & 15.79 & 1.41 & 1.17\\
Star 4 & G5III & 16.64 & 0.84 & 2.02\\
Star 5 & K5III & 14.20 & 2.04 & -0.42\\
MB95019 & G2III & 15.96 & 0.75 & 1.34\\
Star 1 & G5III & 15.66 & 0.96 & 1.04\\
Star 2 & G5III & 17.45 & 1.06 & 2.83\\
Star 3 & G8III & 16.17 & 1.14 & 1.55\\
Star 4 & K3III & 16.66 & 1.18 & 2.04\\
Star 5 & G8III & 16.26 & 1.01 & 1.64\\
MB95030 & M2III & 14.28 & 2.55 & -0.34\\
Star 1 & G8III & 14.80 & 0.54 & 0.28\\
Star 2 & K3I & 12.73 & 1.00 & -1.89\\
Star 3 & K1I & 14.20 & 1.21 & -0.42\\
Star 4 & G5III & 14.55 & 0.57 & -0.07\\

\enddata
\end{deluxetable}


\begin{thebibliography}{}
\bibitem[Alard et al.(1995)]{ala95} Alard, C., Mao, S., Guibert, J. 1995,
\aap, 300, L17
\bibitem[Albrow et al.(1998)]{alb98} Albrow, M., et al. 1998, \apj, 509,
687
\bibitem[Alcock et al.(1997)]{alc97} Alcock, C., et al. 1997, \apj, 491,
436
\bibitem[Bertelli et al.(1994)]{ber94} Bertelli, G., et al. 1994, A\&AS,
106, 275
\bibitem[Butler et al.(1996)]{but96} Butler, R.P., et al. 1996, \pasp,
108, 500
\bibitem[Danks \& Dennefeld(1994)]{dan94} Danks, A.C \& Dennefeld, M.
1994, \pasp, 106, 382
\bibitem[Dravins(1999)]{dra99} Dravins, D. 1999, in Proc. IAU Symp. 170,
``Precise Stellar Radial Velocities'', ed. J.B. Hearnshaw \& C.D.
Scarfe, p.268
\bibitem[Feltzing \& Gilmore(2000)]{fel00} Feltzing, S. \& Gilmore, G.
2000, \aap, 355, 949
\bibitem[Jacoby et al.(1984)]{jac84} Jacoby, G.H., Hunter, D.A.,
Christian, C.A. 1984, \apjs, 56, 257
\bibitem[Kane, 2000]{kan00a} Kane, S.R. 2000, Ph.D. Dissertation,
University of Tasmania, Hobart, Australia
\bibitem[Kane \& Sahu(2000)]{kan00b} Kane, S.R. \& Sahu, K.C. 2000, \apj,
submitted (astro-ph/0011383)
\bibitem[Kiraga \& Paczy\'nski(1994)]{kir94} Kiraga, M. \& Paczy\'nski, B.
1994, \apj, 430, L101
\bibitem[Kurucz(1970)]{kur70} Kurucz, R. 1970, ``Atlas: A computer
program for calculating model stellar atmospheres'', SAO Special Report,
Cambridge: Smithsonian Astrophysical Observatory
\bibitem[Kurucz(1993)]{kur93} Kurucz, R. 1993, ATLAS9 Stellar Atmosphere
Programs and 2 km/s grid. Kurucz CD-ROM No. 13. Cambridge, MA:
Smithsonian Astrophysical Observatory
\bibitem[Mao et al.(1998)]{mao98} Mao, S., Reetz, J., Lennon, D.J. 1998,
\aap, 338, 56
\bibitem[M\'era et al.(1998)]{mer98} M\'era, D., Chabrier, G., Schaeffer,
R. 1998, \aap, 330, 937
\bibitem[Mollerach \& Roulet(1996)]{mol96} Mollerach, S. \& Roulet, E.
1996, \apj, 458, L9
\bibitem[Morgan et al.(1943)]{mor43} Morgan, W.W., Keenan, P.C., Kellman,
E. 1943, ``An Atlas of Stellar Spectra with an Outline of Spectral
Classification'' (University of Chicago Press)
\bibitem[Morse et al.(1991)]{mor91} Morse, J.A., Mathieu, R.D., Levine,
S.E. 1991, \aj, 101, 1495
\bibitem[Paczy\'nski et al.(1994a)]{pac94a} Paczy\'nski, B., et al. 1994a,
\aj, 107, 2060
\bibitem[Paczy\'nski et al.(1994b)]{pac94b} Paczy\'nski, B., et al. 1994b,
\apj, 435, L113
\bibitem[Paczy\'nski \& Stanek(1998)]{pac98} Paczy\'nski, B. \& Stanek,
K.Z. 1998, \apj, 494, L219
\bibitem[Schmidt-Kaler(1982)]{sch82} Schmidt-Kaler, T. 1982, BICDS, 23, 2
\bibitem[Seaton(1979)]{sea79} Seaton, M.J. 1979, \mnras, 187, 73p
\bibitem[Seeds(1999)]{see99} Seeds, M.A. 1999, Foundations of Astronomy
(1999 ed.; Belmont, CA: Wadsworth Pub. Co)
\bibitem[Stanek et al.(1994)]{sta94} Stanek, K.Z., et al. 1994, \apj,
429, L73
\bibitem[Stanek(1995)]{sta95} Stanek, K.Z. 1995, \apj, 441, L29
\bibitem[Stanek et al.(2000)]{sta00} Stanek, K.Z., et al. 2000, Acta
Astron., 50, 191
\bibitem[Terndrup(1988)]{ter88} Terndrup, D.M. 1988, \aj, 96, 884
\bibitem[Tonry \& Davis(1979)]{ton79} Tonry, J., Davis, M. 1979, \aj, 84,
1511
\bibitem[Torres-Dodgen \& Weaver(1993)]{tor93} Torres-Dodgen, A.V. \&
Weaver, W.B. 1993, \pasp, 105, 693
\bibitem[Udalski et al.(1993)]{uda93} Udalski, A., et al. 1993, Acta
Astron., 43, 69
\bibitem[Udry et al.(1999)]{udr99} Udry, S., et al. 1999, in Proc. IAU
Symp. 170, ``Precise Stellar Radial Velocities'', ed. J.B. Hearnshaw
\& C.D. Scarfe, p.383
\bibitem[Walker(1997)]{wal97} Walker, M.A. 1997, \mnras, 287, 629
\bibitem[Wo\'zniak \& Szyma\'nski(1998)]{woz98} Wo\'zniak, P.,
Szyma\'nski, M. 1998, Acta Astron., 48, 269
\bibitem[Zhao et al.(1995)]{zha95} Zhao, H., Spergel, D.N., Rich, R.M.
1995, \apj, 440, L13
\end{thebibliography}
\end{document}